\documentclass[aps,pre,10pt,showpacs,floatfix,onecolumn,nofootinbib,a4paper]{revtex4-2}
\usepackage{amssymb, amsmath, amsthm, mathtools}
\usepackage{bm}
\usepackage{mathrsfs}
\usepackage{hyperref,bookmark}
\usepackage{graphicx}
\usepackage{epsfig,color}
\usepackage{mathrsfs}
\usepackage{verbatim}
\usepackage{url}
\usepackage{subcaption}
\usepackage{float}
\usepackage[greek,english]{babel}
\usepackage{dcolumn}
\usepackage[utf8]{inputenc}

\newtheorem{lemma}{Lemma}

\newcommand{\tr}{\operatorname{tr}}
\newcommand{\sgn}{\operatorname{sgn}}
\newcommand{\arctanh}{\operatorname{arctanh}}
\newcommand{\dd}{\operatorname{d}\!}

\newcommand{\diver}{\operatorname{div}}
\newcommand{\curl}{\operatorname{curl}}

\newcommand{\n}{\bm{n}}
\newcommand{\av}{\bm{a}}

\newcommand{\cv}{\bm{c}}
\newcommand{\nt}{\n_t}
\newcommand{\dotnt}{\dot{\n}_t}
\newcommand{\ddotnt}{\ddot{\n}_t}
\newcommand{\e}{\bm{e}}
\newcommand{\x}{\bm{x}}
\newcommand{\sphere}{\mathbb{S}^2}
\newcommand{\nper}{\bm{n}_\perp}
\newcommand{\normal}{\bm{\nu}}
\newcommand{\body}{\mathscr{B}}
\newcommand{\free}{\mathscr{F}}
\newcommand{\freeF}{\free_\mathrm{F}}
\newcommand{\boundary}{\partial\mathscr{B}}

\newcommand{\vt}{\vartheta}

\newcommand{\framec}{(\e_r,\e_\vt,\e_z)}
\newcommand{\frames}{(\e_r,\e_\vt,\e_\varphi)}
\newcommand{\frameC}{(\e_1,\e_2,\e_3)}

\newcommand{\framed}{(\n_1,\n_2,\n)}

\newcommand{\conf}{\mathsf{S}}

\newcommand{\vae}{\varepsilon}

\newcommand{\nablas}{\nabla\!_\mathrm{s}}
\newcommand{\divs}{\diver\!_\mathrm{s}}

\newcommand{\vv}{\bm{v}}
\newcommand{\w}{\bm{w}}
\newcommand{\bend}{\bm{b}}
\newcommand{\nB}{\bm{n}_{\mathrm{ET}}}
\newcommand{\Wn}{\mathbf{W}(\n)}
\newcommand{\Pn}{\mathbf{P}(\n)}
\newcommand{\Dn}{\mathbf{D}}
\newcommand{\I}{\mathbf{I}}
\newcommand{\twon}{(\n_1,\n_2)}
\newcommand{\zero}{\bm{0}}
\newcommand{\distom}{(S,T,\bend,\Dn)}
\newcommand{\WF}{W_\mathrm{F}}
\newcommand{\grads}{\nabla_{\!\mathrm{s}}}
\newcommand{\area}{A}
\newcommand{\volume}{V}
\newcommand{\adm}{\mathscr{A}}

\begin{document}
\latintext

%\title{Chromonic Liquid Crystals and Second Variation of Frank's Free Energy}
\title{Stability Against the Odds: the Case of Chromonic Liquid Crystals}
\author{Silvia Paparini}
\email{silvia.paparini@unipv.it}
\affiliation{Dipartimento di Matematica, Universit\`a di Pavia, Via Ferrata 5, 27100 Pavia, Italy}
\author{Epifanio G. Virga}
\email{eg.virga@unipv.it}
\affiliation{Dipartimento di Matematica, Universit\`a di Pavia, Via Ferrata 5, 27100 Pavia, Italy}

\date{\today}

\begin{abstract}
The ground state of chromonic liquid crystals, as revealed by a number of recent experiments, is quite different from that of ordinary nematic liquid crystals: it is \emph{twisted} instead of uniform. The common explanation provided for this state within the classical elastic theory of Frank demands that one Ericksen's inequality is violated. Since in general such a violation makes Frank's elastic free-energy functional unbounded below, the question arises as to whether the twisted ground state can be locally stable. We answer this question in the \emph{affirmative}. In reaching this conclusion, a central role is played  by the specific boundary conditions imposed in the experiments on the boundary of rigid containers and by a general formula that we derive here for the second variation of Frank's elastic free energy.	
\end{abstract}

\maketitle

\section{Introduction}\label{sec:intro}
Liquid crystals are anisotropic fluids that fall basically into two broad categories: they are either \emph{thermotropic} or \emph{lyotropic}, depending on whether  it is temperature or concentration, respectively, responsible for driving the formation of these fascinating intermediate phases of soft condensed matter, which are birefringent like crystals and flow like liquids.

Chromonic liquid crystals (CLCs) are lyotropic. They are composed of plank-like molecules with a poly-aromatic core and polar peripheral groups, aggregated in columnar stacks resulting from noncovalent attractions between the poly-aromatic cores.  CLCs are formed by certain dyes, drugs, and short nucleic-acid oligomers in aqueous solutions \cite{dickinson:aggregate,tam-chang:chromonic,mariani:small,zanchetta:phase,nakata:end-to-end,fraccia:liquid}. Their properties are reviewed in the following papers \cite{lydon:chromonic_1998,lydon:handbook,lydon:chromonic_2010,lydon:chromonic,dierking:novel}.

At low concentrations, the supramolecular columns are not long enough to promote the nematic phase, in which their long axes would tend to be organized in a parallel fashion. Upon increasing the concentration, the nematic phase takes eventually over, although the critical concentration turns out to be much lower (see, for example, \cite{nayani:spontaneous}) than that predicted by Onsager's excluded volume theory \cite{onsager:effects}.  
Several unconventional models for the columnar organization, envisioning the possibility that molecular stacks could be either Y-shaped or side-slipped \cite{xiao:structural,park:self-assembly}, have been put forward to try and explain this discrepancy. But this is not the only oddity of these nematic phases.

They do not seem to possess the same ground state as ordinary nematics. When a low-molecular liquid crystal in the nematic phase is left to itself, in the absence of either external disturbing agencies or confining boundaries, the nematic director $\n$, which represents on a macroscopic scale the average orientation of the elongated molecules that constitute the medium, tends to be uniform in space, in a randomly chosen direction. This is \emph{not} what a CLC does.

\begin{figure}[h]
	\centering
\begin{subfigure}[c]{0.45\linewidth}
	\centering
	\includegraphics[width=\linewidth]{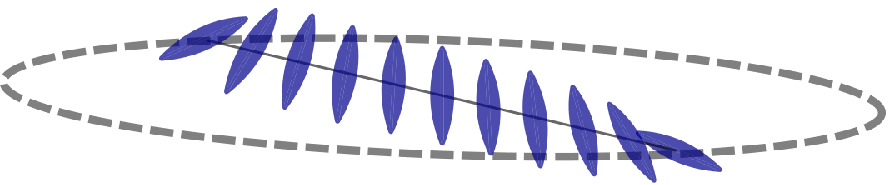}
	%\caption{}
    %\label{fig:f_1}
\end{subfigure}
\quad
\begin{subfigure}{0.45\linewidth}
	\centering
	\includegraphics[width=\linewidth]{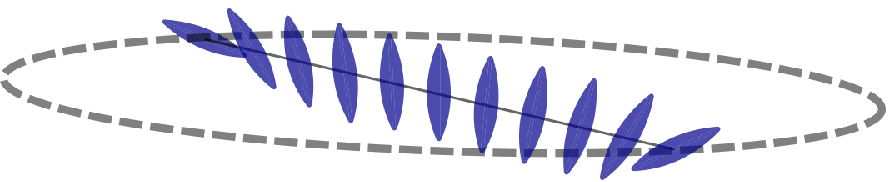}
\end{subfigure}
	\caption{Sketches representing the director arrangement in two symmetric variants of the escaped-twist (ET) distortion within a capillary tube with degenerate planar anchoring conditions on its lateral boundary (schematically indicated by a dashed circle).}
	\label{fig:ET_sketch}
\end{figure}

Experiments in capillary tubes with lateral boundary so smooth as to ensure degenerate planar anchoring for $\n$,\footnote{That is, with $\n$ free to rotate in the local tangent plane.} have shown that the spontaneous distortion is \emph{not} the alignment along the cylinder's axis, which is the only uniform one compatible with the boundary conditions; rather, it is a \emph{twisted} orientation \emph{escaped} along the cylinder's axis, swinging away from it on the cylinder's lateral boundary (see Fig.~\ref{fig:ET_sketch}).
Such an \emph{escaped-twist} (ET) distortion is very similar (but not completely identical to) the \emph{double twist} (DT), which, when energetically  favoured in cholesteric liquid crystals, gives rise to their \emph{blue phases}.\footnote{Blue phases, which owe their name to the colour of their appearance, are exhibited by cholesteric liquid crystals in the proximity of their transition to the isotropic phase. They are attributed to a lattice of line defects that orderly traverse the medium (in different spatial arrangements in the phases labeled BPI and BPII) and to a disordered network in the amorphous phase BPIII (see pp.\,68 and 407 of \cite{kleman:soft} for a quick, but effective introduction to this topic, and \cite{pisljar:blue} for one of the latest most illuminating models). The name \emph{double twist} was coined in \cite{meiboom:theory} (see also \cite{meiboom:lattice} for a fuller presentation of the elastic theory of blue phases), but the spatial arrangement of the director field that gives rise to it had already been precognized in \cite{saupe:molecular}. The difference between ET and DT distortions rests simply on the fact that in the latter the lateral rotation of the director must be tuned appropriately, so that cylindrical tubes with axes at right angles can be glued together generating a network of disclinations.}

ET distortions come with two types of handiness: the director may wind either clockwise or anticlockwise as we progress radially outwards from the cylinder's axis. Being both helicities equally energetic, they are seen with equal probability, and so either singular (point) defects \cite{davidson:chiral} or regular domain walls \cite{nayani:spontaneous} may arise where two ET domains with opposite chirality come together.  

The issue as to whether the ET distortions may or may not embody the ground state of chromonics is tackled in Sect.~\ref{sec:ground}. There, we shall make contact with a notion of elastic  frustration, which arises naturally from the geometric incompatibility of the ET distortions. The latter were first described analytically by Burylov~\cite{burylov:equilibrium}; they exist as solutions to the pertinent Euler-Lagrange equations only when the uniform orientation along the cylinder's axis ceases to be locally stable. We elaborate on this in Sect.~\ref{sec:local}, after having provided in Sect.~\ref{sec:second_variation} a general formula for the second variation of the classical Frank's elastic free-energy functional for nematic liquid crystals. Finally, in Sect.~\ref{sec:conclusions}, we collect our conclusions. The paper is closed by two Appendices: in one, we construct a dynamical analogy for the equilibrium director configurations in a capillary tube and give a phase space representation of the ET distortions; in the other,  we record a number of accessory results that should assist the reader in navigating the main text. 
	
\section{Ground State}\label{sec:ground}
The classical Frank's theory \cite{frank:theory} for low molecular weight nematics has been applied to chromonics as well. Frank's theory is variational in nature; it is based on a celebrated formula\footnote{For which there were antecedents in the works of Zocher~\cite{zocher:effect} and Oseen~\cite{oseen:theory}.} for the elastic free-energy density (per unit volume) which measures the distortional cost $W$ produced by a deviation from a uniform director field $\n$. $W$  is chosen to be the most general frame-indifferent,  even function quadratic in $\nabla\n$, 
\begin{equation}
	\label{eq:free_energy_density}
	W_\mathrm{F}(\n,\nabla\n):=\frac{1}{2}K_{11}\left(\diver\n\right)^2+\frac{1}{2}K_{22}\left(\n\cdot\curl\n\right)^2+ \frac{1}{2}K_{33}|\n\times\curl\n|^{2} + K_{24}\left[\tr(\nabla\n)^{2}-(\diver\n)^{2}\right],
\end{equation}
Here $K_{11}$, $K_{22}$, $K_{33}$, and $K_{24}$ are Frank's elastic constants, material moduli characteristic of each liquid crystal. They are often referred to as the \emph{splay}, \emph{twist}, \emph{bend}, and \emph{saddle-splay} constants, respectively, by the features of the different orientation fields, each with a distortion energy proportional to a single term in \eqref{eq:free_energy_density}. While the elastic modes associated with $K_{11}$, $K_{22}$, and $K_{33}$ can be independently excited in the whole space, albeit not necessarily uniformly (that is, not with the same energy everywhere), the saddle-splay mode driven by $K_{24}$ can only be discerned locally from the other elastic modes.\footnote{One can exhibit an orientation field whose energy density consists only in the $K_{24}$ term just at a point in space (see, for example, Chap.~5 of \cite{virga:variational}).} 

Recently, Selinger~\cite{selinger:interpretation} has reinterpreted the classical Frank's energy \eqref{eq:free_energy_density} by decomposing the saddle-splay mode into a set of other independent modes. The starting point of this decomposition is a novel representation of $\nabla\n$ (see also \cite{machon:umbilic}),
\begin{equation}
	\label{eq:nabla_n_novel}
	\nabla\n=-\bend\otimes\n+\frac12T\Wn+\frac12S\Pn+\Dn,
\end{equation}
where $\bend:=-(\nabla\n)\n=\curl\n\times\n$ is the \emph{bend} vector, $T:=\curl\n\cdot\n$ is the \emph{twist}, $S:=\diver\n$ is the \emph{splay}, $\Wn$ is the skew-symmetric tensor that has $\n$ as axial vector, $\Pn:=\I-\n\otimes\n$ is the projection onto the plane orthogonal to $\n$, and $\Dn$ is a symmetric tensor such that $\Dn\n=\zero$ and $\tr\Dn=0$. By its own definition, $\Dn\neq\zero$ admits the following biaxial representation,
\begin{equation}
	\label{eq:D_representation}
	\Dn=q(\n_1\otimes\n_1-\n_2\otimes\n_2),
\end{equation}
where $q>0$ and $\twon$ is a pair of orthogonal unit vectors in the plane orthogonal to $\n$, oriented so that $\n=\n_1\times\n_2$. It is argued in \cite{selinger:director} that $q$ should be given the name \emph{tetrahedral} splay, to which we would actually prefer \emph{octupolar} splay for the role played by a cubic (octupolar) potential on the unit sphere \cite{pedrini:liquid} in representing all independent \emph{distortion characteristics} $(S,T,b_1,b_2,q)$, but $T$. Here $b_1$ and $b_2$ are the components of the bend vector $\bend$ in the \emph{distortion frame} $\framed$,
\begin{equation}
	\label{eq:bend_components}
	\bend=b_1\n_1+b_2\n_2.
\end{equation} 

By use of the following identity, 
\begin{equation}
	\label{eq:identity}
	2q^2=\tr(\nabla\n)^2+\frac12T^2-\frac12S^2,
\end{equation}
we can easily give Frank's formula \eqref{eq:free_energy_density} the equivalent form
\begin{equation}
	\label{eq:Frank_equivalent}
	W_\mathrm{F}(\n,\nabla\n)=\frac12(K_{11}-K_{24})S^2+\frac12(K_{22}-K_{24})T^2+\frac12K_{33}B^2+2K_{24}q^2,
\end{equation}
where $B^2:=\bend\cdot\bend$. Since $\distom$ are all independent \emph{measures of distortion}, it readily follows from \eqref{eq:Frank_equivalent} that $\WF$ is positive semi-definite whenever
\begin{eqnarray}
	\label{eq:Ericksen_inequalities}
	K_{11}\geqq K_{24}\geqq0,\quad K_{22}\geqq K_{24}\geqq0,\quad K_{33}\geqq0,
\end{eqnarray}
which are the celebrated \emph{Ericksen's inequalities} \cite{ericksen:inequalities}. If these inequalities are satisfied in strict form, the global ground state of $\WF$ is attained on the uniform director field, characterized by
\begin{equation}
	\label{eq:uniform_ground_state}
	S=T=B=q=0.
\end{equation}

More generally, it has been shown \cite{virga:uniform} that besides \eqref{eq:uniform_ground_state} the only \emph{uniform distortions}, that is director fields that fill three-dimensional Euclidean space, having everywhere the same distortion characteristics, are only those for which
\begin{equation}
	\label{eq:uniform_distortions}
	S=0,\quad T=\pm2q,\quad b_1=\pm b_2=b,
\end{equation}
corresponding to Meyers's \emph{heliconical} distortions \cite{meyer:structural} characterizing the ground state of the \emph{twist-bend}
nematic phases identified  experimentally in \cite{cestari:phase}.\footnote{In \eqref{eq:uniform_distortions}, $q$ is \emph{positive} and $b$ arbitrary. As shown in \cite{virga:uniform}, if $q$ vanishes also does $b$ and both forms of uniform distortions reduce to the standard uniform orientation in \eqref{eq:uniform_ground_state}.}

The experimental evidence gathered in \cite{davidson:chiral,nayani:spontaneous} tells us that CLCs within a cylinder with degenerate planar anchoring on the lateral wall acquire either of two oppositely twisted distortions (see Fig.~\ref{fig:ET_sketch}). This shows that the uniform distortion in  \eqref{eq:uniform_ground_state} is \emph{not} the ground state of chromonics, and neither are \eqref{eq:uniform_distortions}, as we shall show, thus entailing a degree of \emph{elastic frustration} in the ground state.

For $\body$ a region in space occupied by the material, the total elastic free energy stored in $\body$ is given by the functional
\begin{equation}
	\label{eq:free_energy}
	\freeF[\n]:=\int_{\body}\WF(\n,\nabla\n)\dd\volume.
\end{equation}
where $\volume$ is the volume measure. It is well-known since the seminal paper of Ericksen~\cite{ericksen:nilpotent}, that the saddle-splay term in $\WF$ plays no role in the minimization of $\freeF$ in the class $\adm(\n_0)$ of admissible distortions for which $\n$ is prescribed on the boundary $\partial\body$,
\begin{equation}
	\label{eq:strong_anchoring}
\n|_{\partial\body}=\n_0. 
\end{equation}
Formally $\adm(\n_0)$ is defined as 
\begin{equation}
	\label{eq:An0}
	\adm(\n_0):=\left\{\n\in\mathcal{H}^1(\body,\mathbb{S}^2): \n_0\in\mathcal{H}^1(\body,\mathbb{S}^2) \hbox{ is the trace of $\n$ on } \boundary\right\},
\end{equation}
where $\mathcal{H}^1$ is the Sobolev space of square integrable functions.
The reason why $K_{24}$ is irrelevant for all $\n\in\adm(\n_0)$ resides in the identity
\begin{equation}\label{eq:null_Lagrangian_identity}
	\tr(\nabla\n)^2-(\diver\n)^2 =\diver\left[(\nabla\n)\n-(\diver\n)\n\right],
\end{equation}
which readily leads us to 
\begin{equation}\label{eq:null_Lagrangian}
	\begin{split}
		\int_\body\left[\tr(\nabla\n)^2-(\diver\n)^2 \right]\dd\volume&=\int_{\boundary}\left[(\nabla\n)\n-(\diver\n)\n\right]\cdot\normal\dd\area\\
		&=\int_{\boundary}\left[(\nablas\n)\n-(\divs\n)\n\right]\cdot\normal\dd\area\\
		&=\int_{\boundary}\left[(\nablas\n_0)\n_0-(\divs\n_0)\n_0\right]\cdot\normal\dd\area,
	\end{split}
\end{equation}
where $\normal$ is the outer unit normal to $\boundary$, $\nablas$ and $\divs$ denote the \emph{surface} gradient and divergence (where only tangential derivatives on $\boundary$ are taken into account), and $\area$ is the area measure.

As a consequence of \eqref{eq:null_Lagrangian}, the saddle-splay energy contributes a constant to $\freeF$, whose value is determined only by the boundary condition $\n_0$, and so it is unable to affect the energy minimizers. This is  why the saddle-splay energy is also called a \emph{null Lagrangian}.

Besides the  \emph{strong anchoring} condition  in \eqref{eq:strong_anchoring}, another boundary condition gives the $K_{24}$-energy  a special form: this is  the \emph{planar degenerate} anchoring, for which, 
\begin{equation}\label{eq:planar_degenerate}
	\n\cdot\normal\equiv0\quad\text{on}\quad\boundary.
\end{equation}
In the latter case, as remarked in \cite{koning:saddle-splay}, the $K_{24}$-integral can be rewritten as 
\begin{equation}\label{eq:K_24_geometric}
	-K_{24}\int_{\boundary}\left(\kappa_1n_1^2+\kappa_2n_2^2\right)\dd\area\,,
\end{equation}
where $\kappa_1$ and $\kappa_2$ are the principal curvatures of $\boundary$, and $n_i$ are the components of $\n$ along the corresponding principal directions of curvature.\footnote{We write the curvature tensor as $\grads\normal=\kappa_1\e_1\otimes\e_1+\kappa_2\e_2\otimes\e_2$, where $\e_1$ and $\e_2$ are unit vectors along the principal directions of curvature of $\boundary$.} It is clear from \eqref{eq:K_24_geometric} that for $K_{24}>0$, which is the strong form of  \eqref{eq:Ericksen_inequalities}, whenever  \eqref{eq:planar_degenerate} applies the saddle-splay energy would locally tend to orient $\n$ on $\boundary$ along the direction of \emph{maximum} (signed) curvature.

For $\body$ a circular cylinder, we now describe the ET distortion that minimizes the free-energy functional $\freeF$ subject to \eqref{eq:planar_degenerate}. Here, we essentially follow Burylov's work \cite{burylov:equilibrium}.\footnote{The same results arrived at in \cite{burylov:equilibrium} were independently reobtained in \cite{davidson:chiral}.} Let $R$ be the radius of the cylinder $\body$ and $L$ its height. We assume that in the frame $\framec$ of cylindrical coordinates $(r,\vt,z)$, with $\e_z$ along the axis of $\body$, $\n$ is represented as
\begin{equation}
	\label{eq:n_bur}
	\n=\sin\beta(r)\e_\vt+\cos\beta(r)\e_z,
\end{equation}
where the \emph{polar} angle $\beta\in[-\pi,\pi]$, which $\n$ makes with the cylinder's axis, depends only on the radial coordinate $r$. Standard computations show that 
\begin{equation}
	\label{eq:grad_B}
	\nabla\n=-\frac{\sin\beta}{r}\e_r\otimes\e_\vt+\cos\beta\beta'\e_\vt\otimes\e_r-\sin\beta\beta'\e_z\otimes\e_r,
\end{equation}
where a prime $'$ denotes differentiation with respect to $r$. As a 
result, the distortion characteristics $(S,T,q)$ of the director field \eqref{eq:n_bur} are given by 
\begin{subequations}\label{eq:distortion_measure_Burylov}
	\begin{align}
		S=&0,\label{eq:S_B}\\
		T=&\beta'+\frac{\cos\beta\sin\beta}{r},\label{eq:T_B}\\
		q=&\frac{1}{2}\left|\beta'-\frac{\cos\beta\sin\beta}{r}\right|,\label{eq:q_B}
	\end{align}
the latter of which follows directly from \eqref{eq:identity}. A direct computation also yields 
\begin{equation}\label{eq:b_B}	
\bend=\frac{1}{r}\sin^2\beta\e_r.
\end{equation}
\end{subequations}
To extract from \eqref{eq:b_B} the remaining distortion characteristics $(b_1,b_2)$, we need to identify the distortion frame $\framed$ of a generic field in \eqref{eq:n_bur}. Letting 
\begin{equation}
	\label{eq:n_perp}
	\nper:=\e_r\times\n=\sin\beta\e_z-\cos\beta\e_{\vt}
\end{equation}
and adopting the representation
\begin{equation}
	\label{eq:n_i_representation}
	\n_1=\cos\alpha\nper+\sin\alpha\e_r,\qquad\n_2=\cos\alpha\e_r-\sin\alpha\nper,
\end{equation}
so that by \eqref{eq:b_B}
\begin{equation}
	\label{eq:b_1_b_2_representation}
	b_1=\frac{1}{r}\sin^2\beta\sin\alpha\quad\text{and}\quad b_2=\frac{1}{r}\sin^2\beta\cos\alpha,
\end{equation}
we readily obtain from \eqref{eq:D_representation} and \eqref{eq:q_B}, with the aid of \eqref{eq:n_perp} and \eqref{eq:n_i_representation}, that
\begin{equation}
	\label{eq:alpha}
	\alpha=\sgn\left(\frac{1}{r}\sin\beta\cos\beta-\beta'\right)\frac{\pi}{4}.
\end{equation}

By changing the variable $r$ into
\begin{equation}\label{eq:rho_definition}
	\rho:=\frac{r}{R},
\end{equation}
which ranges in $[0,1]$, and making use of \eqref{eq:distortion_measure_Burylov} in \eqref{eq:Frank_equivalent}, we arrive at the following reduced functional, $\mathcal{F}[\beta]$, which is an appropriate dimensionless form of Frank's free-energy functional,
\begin{equation}
	\label{eq:free_bur}
	\mathcal{F}[\beta]:=\frac{\freeF[\n]}{2\pi K_{22}L}= \int_0^1\left(\frac{\rho\beta'^2}{2}+\dfrac{1}{2\rho}\cos^2\beta\sin^2\beta+\dfrac{k_3}{2\rho}\sin^4\beta\right) \dd\rho+\frac12(1-2k_{24})\sin^2\beta(1),
\end{equation}
where the boundary term clearly echos\footnote{To be  more precise, an additional boundary term besides the $K_{24}$-energy arises in \eqref{eq:free_bur} from an integration by part of a component of the twist-energy.} \eqref{eq:K_24_geometric}
and the following scaled elastic constants have been introduced,
\begin{equation}
	\label{eq:scaled_elastic_constant}
	k_3 :=\frac{K_{33}}{K_{22}}, \quad k_{24}:=\frac{K_{24}}{K_{22}}, \quad\text{with}\quad K_{22}>0.
\end{equation}
We defer the reader to Appendix~\ref{sec:analogy} for a reinterpretation of  the functional $\mathcal{F}$ in \eqref{eq:free_bur} as the action of an effective dynamical system. For the integral in \eqref{eq:free_bur} to be convergent, $\beta$ must be subject to the condition 
\begin{equation}
	\label{eq:boundary_condition_0}
	\beta(0)=0,
\end{equation}
which amounts to require that $\n$ is along $\e_z$ on the cylinder's axis.\footnote{Actually, the convergence requirement would also be satisfied by enforcing the more general condition $\sin\beta(0)=0$; however, our choice is not restrictive, it rather rests on the nematic symmetry, for which $\n$ and $-\n$ are physically equivalent.}
We seek the functions $\beta=\beta(\rho)$ of class $\mathcal{C}^2$ on $[0,1]$ that satisfy \eqref{eq:boundary_condition_0} and make \eqref{eq:free_bur} stationary.

\begin{figure}[h]
	\centering
	\begin{subfigure}[c]{0.4\linewidth}
		\centering
		\includegraphics[width=\linewidth]{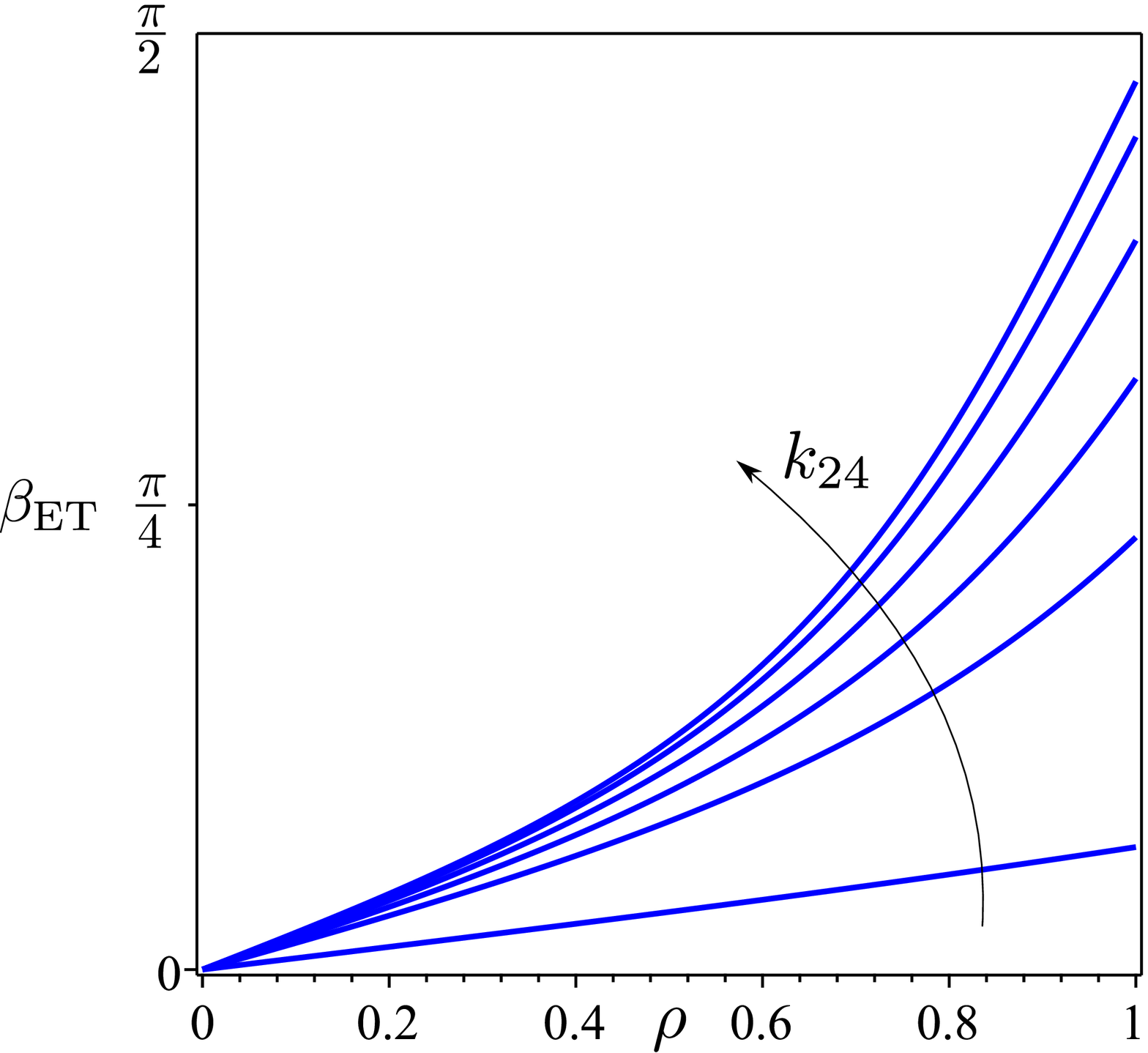}
		\caption{$k_3=10$ and $k_{24}=1.1,\ 2,\ 3,\ 5,\ 10,\ 22.5$.\\For reference, the values measured in \cite{davidson:chiral} correspond to $k_3\approx10$ e $k_{24}\approx22.5$.}
		\label{fig:beta_plot_k24}
	\end{subfigure}
	\qquad
	\begin{subfigure}{0.4\linewidth}
		\centering
		\includegraphics[width=\linewidth]{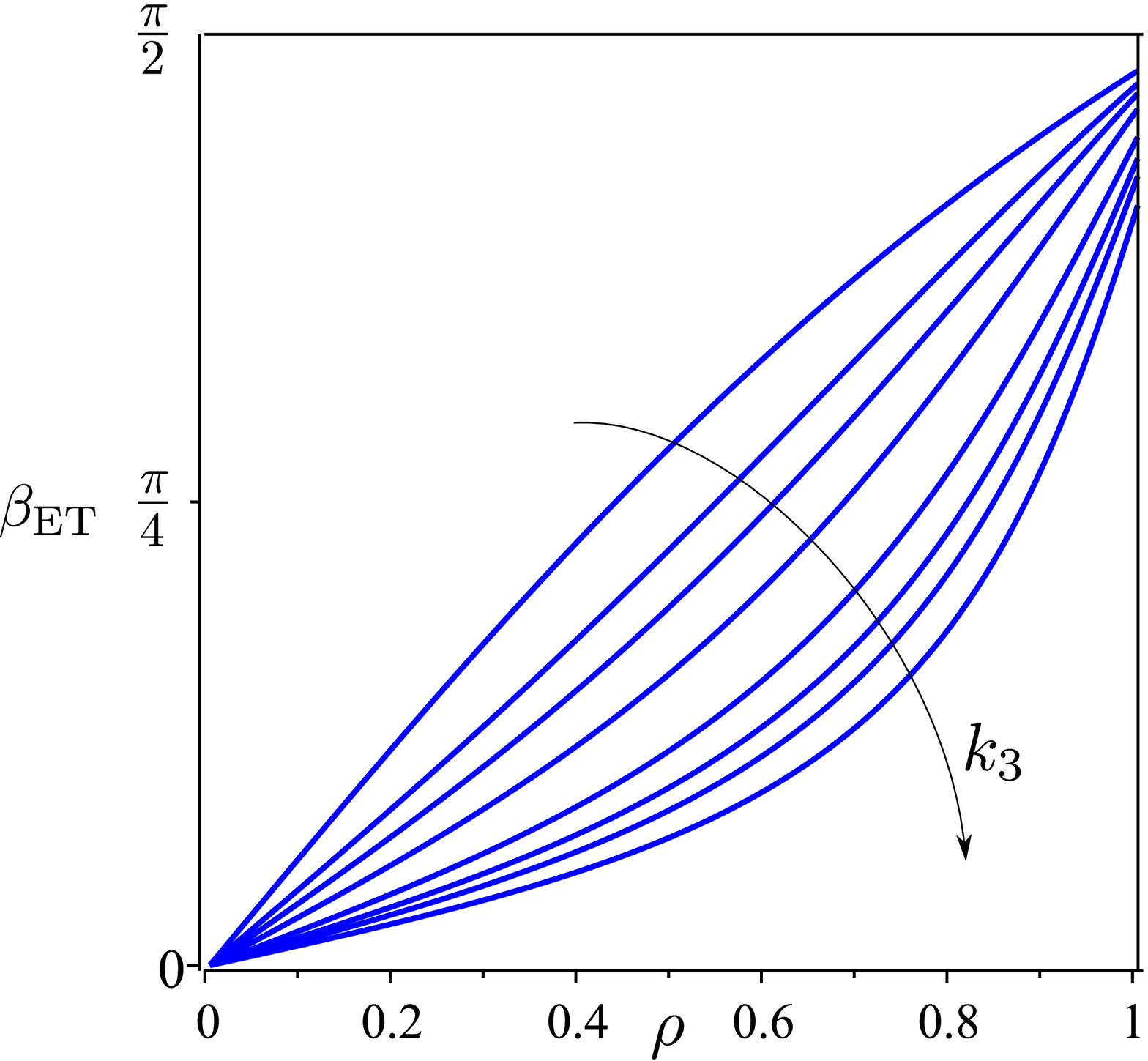}
			\caption{$k_{24}=10$ and $k_3=1,\ 2,\ 3,\ 5,\ 10,\ 20,\ 30$.\\
				For reference, the values measured in \cite{eun:effects} correspond to $k_{3}\approx30$ e $k_{24}\approx7.3$.}
		\label{fig:beta_plot_k3}
	\end{subfigure}
	\caption{Graphs of the function $\beta_\mathrm{ET}$ against $\rho$ for different values of the elastic parameters $k_{24}$ and $k_3$, increasing according to the orientation of the arrow. The twist angle is enhanced for increasing $k_{24}$ and depressed for increasing $k_3$.}
	\label{fig:beta_plots}
\end{figure}

The Euler-Lagrange equations for the functional $\mathcal{F}$ read as
\begin{subequations}\label{eq:burylov_el}
	\begin{align}
		\frac{\cos\beta\sin\beta}{\rho}\left[1+2(k_3-1)\sin^2\beta\right]-\beta'-\rho\beta''&=0, \quad \rho\in(0,1)\label{eq:burylov_el_bulk},\\
		\left.\left[(1-2k_{24})\cos\beta\sin\beta+\beta'\right]\right|_{\rho=1}&=0,\label{eq:burylov_el_sup}
	\end{align}
\end{subequations}
subject to \eqref{eq:boundary_condition_0}. Clearly, the constant 
$\beta\equiv0$ is a trivial solution to these equations. Two other  nonuniform solutions could also exist, depending on the value of $k_{24}$. To show this, we multiply both sides of \eqref{eq:burylov_el_bulk} by $\rho\beta'$ and easily see that this equation has a first integral,
\begin{equation}
	\label{eq:first_intergral_Burylov}
	\left(\rho\beta'\right)^2-\sin^2\beta\left(\cos^2\beta+k_3\sin^2\beta\right)=c,
\end{equation}
where $c$ is an arbitrary constant. The regularity assumption entails that $|\beta'(0)|<\infty$, and so $c$ must vanish for \eqref{eq:boundary_condition_0} to be valid. Two branches of solution thus emanate from $\rho=0$, depending on whether $\beta'(0)$ is positive or negative: they are one opposite to the other and both are obtained by integrating the equation
\begin{equation}
	\label{eq:branches_burylov}
	\beta'=\frac{\sin\beta\cos\beta\sqrt{1+k_3\tan^2\beta}}{\rho}.
\end{equation}
We shall focus on the increasing (positive) branch.  Evaluating \eqref{eq:branches_burylov} for $\rho=1$ and making use of it into the boundary condition \eqref{eq:burylov_el_sup}, we obtain the following equation for $\beta(1)$,
\begin{equation}
	\label{eq:boundary_value_burylov}
	\sin\beta(1)\cos\beta(1)\left[\sqrt{1+k_3\tan^2\beta(1)}+(1-2k_{24})\right]=0,
\end{equation}
which admits a non-vanishing solution 
only for $k_{24}>1$; that is, 
\begin{equation}
	\label{eq:one_value_burylov}
	\beta(1)=\beta_1:=\arctan\left(\frac{2\sqrt{k_{24}(k_{24}-1)}}{\sqrt{k_3}}\right).
\end{equation}
Solving \eqref{eq:branches_burylov} by quadrature yields  
\begin{equation}
	\label{eq:integration_Burylov}
	\ln\rho=\int\frac{\dd \beta}{\sin\beta\cos\beta\sqrt{1+k_3\tan^2\beta}}=\frac12\ln g(\beta)+C,
\end{equation}
where 
\begin{equation}
	\label{eq:g_func_Burylov}
	g(\beta):=\frac{1}{2}-\frac{1}{\sqrt{1+k_3\tan^2\beta}+1}
\end{equation}
and $C$ is an arbitrary constant to be determined so as to satisfy \eqref{eq:one_value_burylov}. With a different constant $C$, we can rewrite \eqref{eq:integration_Burylov} as
\begin{equation}
	\label{eq:C}
	\rho^2=C\left(1-\frac{2}{\sqrt{1+k_3\tan^2\beta}+1}\right),
\end{equation}
which is valid for all $\rho\in[0,1]$ only if $C>1$. By requiring that \eqref{eq:C} agrees with \eqref{eq:one_value_burylov}, we determine $C$ and arrive at the following solution
\begin{equation}
	\label{eq:bur_solution}
	 \beta_\mathrm{ET}(\rho):=\arctan\left(\frac{2\sqrt{k_{24}(k_{24}-1)}\rho}{\sqrt{k_{3}}\left[k_{24}-(k_{24}-1)\rho^2\right]}\right), 
\end{equation}
which together with its opposite\footnote{That is, the  decreasing solution branch of \eqref{eq:branches_burylov}.} $- \beta_\mathrm{ET}$ represent the two variants of the ET distortion (with opposite chiralities). The function $\beta_\mathrm{ET}$ is plotted against $\rho$ in Fig.~\ref{fig:beta_plots}, which illustrates the antagonistic roles played by the elastic parameters $k_{24}$ and $k_3$.

By use of both \eqref{eq:first_intergral_Burylov} and \eqref{eq:integration_Burylov} in \eqref{eq:free_bur}, we express $\mathcal{F}[\beta_\mathrm{ET}]$ in the form,
\begin{equation}
	\label{eq:free_energy_ET_beta_1}
	\mathcal{F}_\mathrm{ET}:=\mathcal{F}[\beta_\mathrm{ET}]=\int_0^{\beta_1}\sin\beta\sqrt{\cos^2\beta+k_3\sin^2\beta}\dd\beta+\frac12(1-k_{24})\sin^2\beta_1,
\end{equation} 
which by \eqref{eq:one_value_burylov} can be written explicitly in terms of the reduced elastic constants only,
\begin{equation}
	\label{eq:ET_free_energy}
	\mathcal{F}_\mathrm{ET}=
	\begin{cases}
		1-k_{24}+\frac{1}{2}\frac{k_3}{\sqrt{1-k_3}}\arctanh\left(\frac{2\sqrt{1-k_3}(k_{24}-1)}{k_3+2(k_{24}-1)}\right), \quad&k_3\leqq1,\\ 1-k_{24}+\frac{1}{2}\frac{k_3}{\sqrt{k_3-1}}\arctan\left(\frac{2\sqrt{k_3-1}(k_{24}-1)}{k_3+2(k_{24}-1)}\right),\quad&k_3\geqq1,
	\end{cases}
\end{equation}
an expression that, for $k_{24}>1$, can be shown to be negative in both instances, as the following inequalities hold true,
\begin{align}
	\label{eq:burylov_energy_inequalities}
1-k_{24}\leqq\mathcal{F}_\mathrm{ET}\leqq-\frac{2(k_{24}-1)^2}{2k_{24}-1}&<0,\quad\text{for}\quad0\leqq k_3\leqq1,\\
-\frac{2(k_{24}-1)^2}{2k_{24}-1}\leqq\mathcal{F}_\mathrm{ET}&<0,\quad\text{for}\quad k_3\geqq1.
\end{align}
$\mathcal{F}_\mathrm{ET}$, as given by \eqref{eq:ET_free_energy} as a function of $k_3$, is continuous along with its derivatives at $k_3=1$.\footnote{Apart from a different scaling of the constant $K_{24}$, the formula in \eqref{eq:ET_free_energy} for $k_3\geqq1$ coincides with equation (5) of \cite{davidson:chiral}. } Moreover, the same formula is also valid for the energy of the mirror image $-\beta_\mathrm{ET}$ of $\beta_\mathrm{ET}$. Thus, whenever the ET distortion is permitted, that is, for $k_{24}>1$, it possesses less elastic free energy than the uniform alignment $\n\equiv\e_z$, and so it becomes eligible for the ground state of CLCs, at least within the cylindrical confinement investigated experimentally.

The price to pay to model mathematically the experimental observations with the ET distortion is to renounce one of Ericksen's inequalities, thus accepting that Frank's functional in \eqref{eq:free_energy} may be unbounded below, jeopardizing in general its coercivity. This has no noxious consequences whenever the degenerate boundary condition \eqref{eq:planar_degenerate} is prescribed, as the integral in \eqref{eq:K_24_geometric} is bounded below and the remaining terms in \eqref{eq:free_energy_density} are well-behaved, provided that the constants $K_{11}$, $K_{22}$, and $K_{33}$ are all positive. Instances are known in the literature where appropriate boundary conditions salvage a functional that in other, more general circumstances would fail to attain its minimum (see, for example, \cite{day:sphere}). Here a similar situation arises with the complicity of cylindrical symmetry. Fearing, however, a potentially latent pathology emerging from the lack of boundedness in the general setting, we find it especially appropriate to study the local stability of the ET distortion, albeit against perturbations that do not break cylindrical symmetry. 

To this end in the following section we shall derive a general formula for the second variation of $\freeF$. Before that, we pause to illuminate further the ET field in terms of the distortion characteristics. By use of \eqref{eq:branches_burylov} in \eqref{eq:distortion_measure_Burylov}, \eqref{eq:b_1_b_2_representation}, and \eqref{eq:alpha}, we see that (for the positive branch) 
\begin{subequations}\label{eq:characteristics_burylov}
\begin{align}
T&=\frac{1}{r}\sin\beta\cos\beta(\sqrt{1+\tan^2\beta}+1),\\
q&=\frac{1}{2r}|\sin\beta|\cos\beta	(\sqrt{1+\tan^2\beta}-1),\\
\alpha&=-\frac{\pi}{4},\quad b_1=-\frac{1}{\sqrt{2}r}\sin^2\beta=-b_2,
\end{align} 
\end{subequations}  
where $\beta=\beta_\mathrm{ET}(\rho)$ as in \eqref{eq:bur_solution} (with $r=\rho R$), whereas for $\beta=-\beta_\mathrm{ET}(\rho)$ (the negative branch), $T$ changes its sign, as do $\alpha$, $b_1$, and $b_2$, while $q$ remains unchanged. Although the characteristics in \eqref{eq:characteristics_burylov} are all different from zero, in the limit as $r\to0$ they yield
\begin{equation}
	\label{eq:distortion_measure_Burylov_rho_0}
	S=q=b_1=b_2=0, \quad T=\frac{4\sqrt{k_{24}-1}}{R\sqrt{k_3k_{24}}},
\end{equation}
suggesting that each ET distortion has a single non-vanishing characteristic only along the cylinder's axis. Being such a single distortion mode unable to fill space, it can be accommodated within the cylinder only by igniting all other modes. We may stress the non-uniformity of ET distortions by saying that they represent a \emph{pseudo} ground state of CLCs, one that is bound to change from place to place in space. Alternatively, we may interpret their lack of uniformity as a result of elastic \emph{frustration} induced by a cylindrical confinement, whose geometric measure transpires in \eqref{eq:distortion_measure_Burylov_rho_0} through the radius $R$. In the following, we shall regard both ways of saying as semantic variants of the same physical notion.

\section{Second Variation}\label{sec:second_variation}
To compute the second variation of Frank's elastic free-energy functional, we devise here a variant to the classical method: we represent a generic perturbation of a given director field $\n$ via a path $\nt(\x)$ on the unit sphere $\sphere$ parameterized in $t\in(-\vae,\vae)$ and going through the point $\n(\x)$.
The mapping $t\mapsto\nt(\x)$ defines for each $\x\in\body$ a trajectory in $\mathbb{S}^2$, which for $t=0$ satisfies $\n_0(\x)=\n(\x)$. Thus, as the virtual time $t$ spans $(-\vae,\vae)$, the vector fields $\nt$ on $\body$ are instantaneous realizations of diverse perturbations of $\n$. 
The constraint of unimodularity for $\nt$ must be valid for all $t$, and so,
\begin{equation}
\label{eq:diff_unimodularity}
\dotnt\cdot\nt=0,
\end{equation}
where a superimposed dot denotes differentiation with respect to $t$. We define the field 
\begin{equation}
\label{eq:vv}
\vv:=\dotnt|_{t = 0},
\end{equation}
which by \eqref{eq:diff_unimodularity}
is orthogonal to $\n$,
\begin{equation}
	\label{eq:v_dot_n=0}
	\vv\cdot\n=0.
\end{equation}
By further differentiating both sides of \eqref{eq:diff_unimodularity} with respect to $t$, we arrive at
\begin{equation}
\label{eq:diff_diff_unimodularity}
\ddotnt\cdot\nt+\dotnt\cdot\dotnt=0.
\end{equation}
The field $\ddotnt|_{t=0}$ represents a second-order perturbation for $\n$. It follows from \eqref{eq:diff_diff_unimodularity} and \eqref{eq:vv} that 
\begin{equation}
\label{eq:ddotn}
\ddotnt|_{t=0}=\w-v^2\n,
\end{equation}
where $\w$, precisely like $\vv$, is an
arbitrary vector field orthogonal to $\n$, and $v^2=\vv\cdot\vv$. We say that $\vv $ and $\w$ are the first- and second-order \emph{outer variations} of $\n$. 

The \emph{first variation} of $\freeF$ at the field $\n$ is a linear functional of $\vv$ defined as
\begin{equation}
\label{eq:free_diff}
\delta\freeF(\n) [\vv] :=\frac{\dd}{\dd t}\freeF[\nt]|_{t=0}=\int_\body\dot{W}_\mathrm{F}\left(\nt,\nabla\nt\right)|_{t=0}\dd V.
\end{equation}
An equilibrium director field $\n$ makes $\delta\freeF(\n)[\vv]$ vanish for all $\vv$ satisfying \eqref{eq:v_dot_n=0}. With the aid of \eqref{eq:vv} and \eqref{eq:trace_n_second} we obtain that
\begin{align}
\label{eq:weak_euler_lagrange}
\delta\freeF(\n)[\vv]:=\int_{\body}&\Big\{K_{11}\diver\n\diver\vv+K_{22}\n\cdot\curl\n\cdot\left(\vv\cdot\curl\n+\n\cdot\curl\vv\right)\nonumber\\
&+K_{33}\left(\n\times\curl\n\right)\cdot\left(\vv\times\curl\n+\n\times\curl\vv\right) %\nonumber \\&
+2K_{24}\left(\tr(\nabla\n\nabla\vv)-\diver\n\diver\vv\right)\Big\}\dd V
\end{align}
with $\vv $ sufficiently regular.\footnote{The reader is deferred to Appendix~\ref{sec:identities} for a number of ancillary results used here.}

Similarly, the \emph{second variation} of $\freeF$ at the field $\n$ is the functional defined by  
\begin{equation}
\label{eq:free_diff_diff}
\delta^2\freeF(\n) [\vv,\w] :=\frac{\dd^2}{\dd t^2}\free[\nt]|_{t=0}=\int_\body\ddot{W}_\mathrm{F}\left(\nt,\nabla\nt\right)|_{t=0}\dd V,
\end{equation}
which is linear in $\w$ and quadratic in $\vv$.
By use of \eqref{eq:ddotn} and \eqref{eq:second_derivative_W}, we give \eqref{eq:free_diff_diff} the following form 
\begin{align}
\label{eq:second_variation}
\delta^2\freeF(\n)[\vv,\w]=\delta\freeF(\n)[\w]+\int_{\body}&\Big\{\left(K_{11}-2K_{24}\right)\left[\left(\diver\vv\right)^2-v^2\left(\diver\n\right)^2-\left(\diver\n\right)\n\cdot\nabla v^2\right]+\nonumber\\
&+K_{22}\Big[(\vv\cdot\curl\n+\n\cdot\curl\vv)^2\nonumber\\
&+2(\n\cdot\curl\n)(\vv\cdot\curl\vv-v^2\n\cdot\curl\n)\Big]\nonumber\\
&+K_{33}\Big[\left|\vv\times\curl\n+\n\times\curl\vv\right|^2\nonumber\\
&+(\n\times\curl\n)\cdot(\vv\times\curl\vv-2v^2\n\times\curl\n-\nabla v^2)\Big]\nonumber\\
&+2K_{24}\left[\tr\left(\nabla\vv\right)^2-v^2\tr\left(\nabla\n\right)^2+\n\times\curl\n\cdot\nabla v^2\right]\Big\}\dd V,
\end{align}
which reduces to only its quadratic component in $\vv$ whenever $\n$ is an equilibrium field. In this latter case, with a slight abuse of notation, we shall simply set $\delta^2\freeF(\n)[\vv]:=\delta^2\freeF(\n)[\vv,\zero]$.

\subsection{Simple Applications}
To put to the test the formula for the second variation of $\freeF$ in \eqref{eq:second_variation}, here we compute it for two special equilibrium fields.\footnote{Which are actually \emph{universal} equilibrium fields, as they solve the equilibrium equations for any frame-indifferent elastic free-energy density $W(\n,\nabla\n)$, as shown in  \cite{ericksen:general}.}
\subsubsection{Uniform Field}\label{sec:uniform_solution}
We begin with a director  field $\n$ uniformly constant in space and see how its stability is related to Ericksen's inequalities \eqref{eq:Ericksen_inequalities}. With no loss of generality, for a given Cartesian frame $\frameC$, we apply  \eqref{eq:second_variation} to the  director field $\n\equiv\e_3$, 
\begin{equation}
\label{eq:second_variation_natural}
\delta^2\freeF(\e_3)[\vv]=\int_{\body}\left\{\left(K_{11}-2K_{24}\right)\left(\diver\vv\right)^2+K_{22}\left(\n\cdot\curl\vv\right)^2+K_{33}\left|\n\times\curl\vv\right|^2+2K_{24}\tr\left(\nabla\vv\right)^2\right\}\dd V,
\end{equation}
where $\vv\cdot\e_3\equiv0$. It follows from this latter identity that  
\begin{equation}
\label{eq:gradient_natural}
\nabla\vv:=\e_1\otimes\av+\e_2\otimes \cv,
\end{equation}
with $\av$ and $\cv$ arbitrary vector fields. Letting $a_i$ and $c_i$ be the components of $\av$ and $\cv$ in the frame $\frameC$, we easily compute
\begin{subequations}\label{eq_distorsion_measures_natural}
\begin{align}
\diver\vv&=a_1+c_2,\label{eq:div_uniform}\\
\e_3\cdot\curl\vv&=c_1-a_2,\label{eq:twist_uniform}\\
\e_3\times\curl\vv&=-a_3\e_1-c_3\e_2,\label{eq:bend_uniform}\\
\tr\left(\nabla\vv\right)^2&=a_1^2+2a_2c_1+c_2^2,\label{eq:tr_uniform}
\end{align}
\end{subequations}
which give \eqref{eq:second_variation_natural} the following form,
\begin{equation}
\label{eq:second_variation_natural_fin}
\delta^2\freeF(\e_3)[\vv]=\int_{\body}\left\{K_{11}\left(a_1^2+c_2^2\right)+2\left(K_{11}-2K_{24}\right)a_1b_2+
K_{22}\left(a_2^2+c_1^2\right)-2(K_{22}-2K_{24})a_2c_1+K_{33}\left(a_3^2+c_3^2\right)\right\}\dd V.
\end{equation}
For a general domain $\body$ with no boundary conditions for $\vv$, the integrand on the right-hand side of \eqref{eq:second_variation_natural_fin} is the sum of three independent quadratic forms, so  that $\delta^2\free(\e_3)[\vv_0]$ is positive whenever all these forms are not negative, which is precisely when  Ericksen's inequalities \eqref{eq:Ericksen_inequalities} are satisfied. 

\subsubsection{Radial Hedgehog}
The radial hedgehog is the director field that in the frame $\frames$ of spherical coordinates is represented as $\n=\e_r$. Standard computations give $\curl\e_r=\zero$,  $\diver\e_r = 2/r$, and $\tr(\nabla\e_r)^2=2/r^2$, so that \eqref{eq:second_variation} reduces to
\begin{equation}
\begin{split}
\label{eq:second_variation_hedgehog}
\delta^2\freeF(\e_r)[\vv]=\int_\body\Big\{\left(K_{11}-2K_{24}\right)\left[\left(\diver\vv\right)^2-\frac{2}{r^2}v^2\right]+K_{22}\left(\e_r\cdot\curl\vv\right)\\+
K_{33}\left|\n\times\curl\vv\right|^2
+2K_{24}\left[\tr\left(\nabla\vv\right)^2-\frac{2}{r^2}v^2\right]\Big\}\dd V.
\end{split}
\end{equation}
For $\body$ a sphere with center in the origin of the frame $\frames$ and under the strong anchoring condition $\vv|_{\partial\body}=\zero$, this formula coincides with equation (2.3) of \cite{kinderlehrer:second}, where
the second variation \eqref{eq:second_variation_hedgehog} was proven to be positive in the class of perturbations $\vv\in\mathcal{H}^1\left(B,\mathbb{R}^3\right)\cap L^{\infty}\left(B,\mathbb{R}^3\right)$ with compact support and satisfying $\vv\cdot\e_r=0$, whenever the following inequality is satisfied,
\begin{equation}
\label{eq:inequalities_hedgehog}
8\left(K_{22}-K_{11}\right)+K_{33}\geq0.
\end{equation}

\section{Local Stability}\label{sec:local}
In this section, we study the local stability of the ET field $\nB$, delivered by \eqref{eq:n_bur} when $\beta=\beta_\mathrm{EF}$ as in \eqref{eq:bur_solution}. We shall focus our attention on this chiral variant of the ET field, as that with opposite chirality, represented by $\beta=-\beta_\mathrm{ET}$, has by symmetry the same energy and the same stability character. 

In the scaled radial coordinate $\rho$ introduced in \eqref{eq:rho_definition}, the perturbation field $\vv$, subject to $\nB\cdot\vv=0$, is represented in the frame $\framec$ of cylindrical coordinates as
 \begin{equation}
\label{eq:Burylov_perturbation}
\vv:=f(\rho)\e_r-g(\rho)\cos\beta_\mathrm{ET}(\rho)\e_\vt+g(\rho)\sin\beta_\mathrm{ET}(\rho)\e_z,
\end{equation}
where $f$ and $g$ are absolutely continuous functions on $[0,1]$. The validity of the degenerate planar anchoring condition \eqref{eq:planar_degenerate} for the perturbed director field requires that 
\begin{equation}
\label{eq:f_R}
f(1)=0,
\end{equation}
while $g(1)$ remains completely arbitrary. Moreover, for the regularity of $\vv$ along the cylinder's axis (and the integrability of the perturbed energy), we also request that
\begin{equation}
\label{eq:f_g_0}
f(0)=g(0)=0. 
\end{equation}

Making use in \eqref{eq:second_variation} of the identities \eqref{eq:distortion_measures_nB_v} recorded in Appendix~\ref{sec:identities}, 
we arrive at the following  dimensionless form of the second variation $\delta^2\freeF(\nB)[\vv]$ in the class of perturbation fields described by  \eqref{eq:Burylov_perturbation},
\begin{align}
\label{eq:second_variation_burylov}
\delta^2\mathcal{F}[f,g]&:=\frac{\delta^2\freeF(\nB)[\vv]}{2\pi L K_{22}}\nonumber \\
&=2\int_0^1\left\{\left[-2\cos^2\beta\sin^2\beta+\frac{\left(-\sin^2\beta+\cos^2\beta\right)^2}{2}+k_3\left(-\sin^2\beta+3\cos^2\beta\right)\sin^2\beta\right]\frac{g^2}{\rho}+\frac{\rho g'^2}{2}\right\}\dd \rho\nonumber\\
&+\left(1-2k_{24}\right)\left(-\sin^2\beta+\cos^2\beta\right)g^2(1)\nonumber\\
&+\int_0^1\left\{\left[k_1-2\left(\rho\beta'+\cos\beta\sin\beta\right)^2+k_3\left(\rho^2\beta'^2+\sin^2\beta+4\cos\beta\sin\beta\beta'-2\sin^4\beta\right)\right]\frac{f^2}{\rho}+ k_1 \rho f'^2\right\}\dd \rho,
\end{align}
where
\begin{equation}
	\label{eq:k_1_definition}
	k_1:=\frac{K_{11}}{K_{22}},
	\end{equation}
while $k_3$ and $k_{24}$ are as in \eqref{eq:scaled_elastic_constant}, and $\beta=\beta_\mathrm{EF}$ as in \eqref{eq:bur_solution}.
Integrations by parts  have been performed in \eqref{eq:second_variation_burylov} by employing both \eqref{eq:f_R} and \eqref{eq:f_g_0}.

In \eqref{eq:second_variation_burylov} the scalar perturbations $f$ and $g$ are decoupled and $\delta^2\mathcal{F}[f,g]$ can be regarded as the sum of two independent functionals. When $f=0$ and $g\neq0$, we say that $\vv$ is an \emph{azimuthal} perturbation. On the contrary, when $f\neq0$ and $g=0$, we say that $\vv$ is a \emph{radial} perturbation. We shall consider these perturbations separately below; the local stability of the field $\nB$ requires that $\delta^2\mathcal{F}[f,g]$ be positive for both types of perturbation.

Our study will be limited to the case $k_3\geqq1$, which according to the experimental measurements of \cite{zhou:elasticity_2012} and \cite{zhou:elasticity_2014} is the only relevant to chromonics, the materials of interest to us in this work.

\subsection{Azimuthal Perturbations}\label{sec:azimuthal_perturbations}
To compute $\delta^2\mathcal{F}[0,g]$, we take advantage of the fact that $\beta_\mathrm{ET}$ is a monotonically increasing function of $\rho$, ranging in the interval $[0,\beta_1]$, with $\beta_1<\frac{\pi}{2}$ as in \eqref{eq:one_value_burylov}, as $\rho$ ranges in $[0,1]$. This allows us to express $g$ as a function of $\beta$, and ultimately as a function, $U(u)$, of the variable
\begin{equation}
	u:=\frac{\sin\beta}{A},
\end{equation} 
which ranges in $[0,1]$, as $A$ is defined by 
\begin{equation}
	\label{eq:A_func}
	A:=\sin\beta_1=\frac{2\sqrt{k_{24}(k_{24}-1)}}{\sqrt{k_3+4k_{24}(k_{24}-1)}}.
\end{equation}
This cascade of changes of variables is characterized by the following relations involving their differentials,
\begin{equation}
	\label{eq:differentials}
	\frac{\dd\rho}{\rho}=\frac{\dd\beta}{\sin\beta\sqrt{\cos^2\beta+k_3\sin^2\beta}}=\frac{\dd u}{u\sqrt{1-(Au)^2}\sqrt{1+(k_3-1)(Au)^2}}.
\end{equation}
Their use gives $\delta^2\mathcal{F}$ the following, more compact form,
\begin{equation}
	\label{eq:second_variation_az_k3A}
	\delta^2\mathcal{F}[U]=\int_0^{1}\left[\frac{\phi(k_3,A,u)}{\gamma(k_3,A,u)}U(u)^2+\gamma(k_3,A,u) U'(u)^2\right]\dd u+q_0(k_3,A)U(1)^2,
\end{equation}
where a prime now denotes differentiation with respect to $u$, we have set
\begin{subequations}\label{eq:phi_gamma_q0}
	\begin{align}
		\phi(k_3,A,u):=&\left(1-2(Au)^2\right)\left(1+2(k_3-1)(Au)^2\right) +4(k_3-1)(1-(Au)^2)(Au)^2, \label{eq:phi_func}\\
		\gamma(k_3,A,u):=&u\sqrt{1-(Au)^2}\sqrt{1+(k_3-1)(Au)^2}, \label{eq:gamma_func}\\
		q_0(k_3,A):=&-\frac{(1-2A^2)\sqrt{1+(k_3-1)A^2}}{\sqrt{1-A^2}},\label{eq:q_0func}
	\end{align}
\end{subequations}
and $U$ belongs to the admissible class
\begin{equation}
	\label{eq:class_thetaz_U}
	\mathcal{A}:= \left\{U \in AC[0,1]: \ U(0)=0\right\}. 
\end{equation}

The elastic constants enter \eqref{eq:second_variation_az_k3A} through the pair of dimensionless parameters $(k_3,A)$. It follows from \eqref{eq:A_func} that $A$ ranges in $(0,1)$ for all $k_{24}\in(1,\infty)$ and $k_3\in[1,\infty)$. Conversely, $k_{24}$ is expressed in terms of $(k_3,A)$ by
\begin{equation}
	\label{eq:k_4_func}
	k_{24}=\frac{\sqrt{1-A^2}+\sqrt{1+(k_3-1)A^2}}{2\sqrt{1-A^2}},
\end{equation}
so that the correspondence between the half-space $\mathsf{H}:=\{(k_3,k_{24}):k_3\in[1,\infty), k_{24}\in(1,\infty)\}$ and the strip $\mathsf{S}:=\{(k_3,A):k_3\in[1,\infty), k_{24}\in(1,\infty)\}$ is one-to-one as the Jacobian determinant of the transformation is positive,
\begin{equation}
	\label{eq:jacobian_A_k3}
	\frac{\partial(k_3,k_{24})}{\partial(k_3,A)}=\frac{k_3}{2}\frac{A}{(1-A^2)^{3/2}\sqrt{1+(k_3-1)A^2}}>0.
\end{equation}

Here we show that for all $(k_3,A)\in\conf$ the functional in \eqref{eq:second_variation_az_k3A} is positive definite in the class $\mathcal{A}$. Our desired conclusion will be reached in two steps. First, we see that the functions
\begin{equation}
	\label{eq:lemma_li_application}
	\widetilde{U}:=\sqrt{\gamma}U, \quad Q(u,\widetilde U):=1, \quad G(u,\widetilde U):=\frac{3}{4}\frac{\gamma'}{\gamma}\widetilde{U}^2=\frac{3}{4}\gamma'U^2,
\end{equation}
obey all three conditions of Lemma \ref{lemma:li1995} in Appendix~\ref{sec:identities} for  $U\in \mathcal{A}$ and $\gamma$  as in \eqref{eq:gamma_func}. Thus, it follows from  \eqref{eq:energy_comparison} that
\begin{equation}
	\label{eq:sqrtgamma_U_prime}
	\int_0^1\left[\left(\sqrt{\gamma}U\right)'\right]^2 \dd u \geqq -\int_0^1\left(\frac{3}{4}\frac{\gamma'^2}{\gamma}+\frac{3}{2}\gamma''\right) U^2 \dd U+\frac{3}{2}\gamma'|_{u=1}U(1)^2.
\end{equation}
Second, from the identity
\begin{equation}
	\label{eq:diff_sqrt_gamma_U}
	\left[\sqrt{\gamma}U\right]'=\frac{\gamma'}{2\sqrt{\gamma}}U+\sqrt{\gamma}U',
\end{equation}
we readily obtain that
\begin{equation}
	\label{eq:gamma_U_prime}
	\int_0^1\gamma U'^2\dd u=\int_0^1\left\{\left[\left(\sqrt{\gamma}U\right)'\right]^2-\frac{\gamma'^2}{4\gamma}U^2-\gamma'U U'\right\}\dd u.
\end{equation}
Making use of \eqref{eq:sqrtgamma_U_prime} in \eqref{eq:gamma_U_prime} and integrating by parts, we finally arrive at
\begin{equation}
	\label{eq:gamma_U_prime_final}
	\int_0^1\gamma U'^2\dd u\geqq\int_0^1\left(-\frac{\gamma'^2}{\gamma}-\gamma''\right)U^2\dd u+\gamma'|_{u=1}U(1)^2.
\end{equation}
We are now in a position to conclude our proof, as combining \eqref{eq:gamma_U_prime_final} with \eqref{eq:second_variation_az_k3A} we estimate
\begin{equation}
	\label{eq:bound_from_below_az}
	\delta^2\mathcal{F}[U]\geqq\int_0^{1}\left(\frac{\phi}{\gamma}-\frac{\gamma'^2}{\gamma}-\gamma''\right)U(u)^2\dd u+\left(q_0+\gamma'|_{u=1}\right)U(1)^2,
\end{equation}
and so, for $\delta^2\mathcal{F}$ to be positive on $\mathcal{A}$ it suffices that the following inequalities are valid
\begin{equation}
	\label{eq:condition_stab}
	\frac{\phi}{\gamma}-\frac{\gamma'^2}{\gamma}-\gamma''\geqq 0, \qquad q_0+\gamma'|_{u=1}\geqq0,
\end{equation}
which by \eqref{eq:phi_gamma_q0} reduce to
\begin{equation}
	\label{eq:final_conditions}
	\frac{7A^2u}{\sqrt{1+(k_3-1)(Au)^2}\sqrt{1-(Au)^2}}\left[\frac{4}{7}+(k_3-1)(Au)^2\right]\geqq 0, \quad \frac{A^2(1-A^2)}{\sqrt{1+(k_3-1)(Au)^2}\sqrt{1-(Au)^2}}(k_3-1)\geqq 0,
\end{equation}
and are easily seen to be valid for all $u\in[0,1]$ and $(k_3,A)\in\conf$.

\subsection{Radial Perturbations}\label{sec:radial_perturbations}
Now, we ignite the radial modes in the second variation $\delta^2\mathcal{F}$ in \eqref{eq:second_variation_burylov}, while silencing the azimuthal ones,
\begin{align}
\label{eq:second_variation_burylov_radial_rho}
\delta^2\mathcal{F}[f,0]=\int_0^1&\left\{\left[k_1-2\left(\rho\beta'+\cos\beta\sin\beta\right)^2+k_3\left(\rho^2\beta'^2+\sin^2\beta+4\cos\beta\sin\beta\beta'-2\sin^4\beta\right)\right]\frac{f^2}{\rho} + k_1 \rho f'^2\right\}\dd \rho,
\end{align}
where $\beta=\beta_\mathrm{ET}$ and $f$ is subject to \eqref{eq:f_g_0} and \eqref{eq:f_R}. Here, as suggested by the experimental evidence presented in \cite{zhou:elasticity_2012,zhou:elasticity_2014}, we assume that for chromonics $k_1\geqq1$. This inequality, combined with the same changes of variables performed in Sect.~\ref{sec:azimuthal_perturbations}, leads us to replace \eqref{eq:second_variation_az_k3A} with
\begin{equation}
\label{eq:second_variation_r_k3A}
\delta^2\mathcal{F}[U]\geqq\int_0^{1}\left[\frac{\psi(k_3,A,u)}{\gamma(k_3,A,u)}U(u)^2+\gamma(k_3,A,u) U'(u)^2\right]\dd u,
\end{equation}
where now 
\begin{equation}
	\label{eq:phi_func_r}
	\psi(k_3,A,u):=1+2(k_3-2)(Au)^2+(k_3-4)(k_3-1)(Au)^4+4(k_3-1)(Au)^2\sqrt{1-(Au)^2}\sqrt{1+(k_3-1)(Au)^2}
\end{equation}
and $U\in\mathcal{A}$ is subject to the additional condition that $U(1)=0$.

Following  the same lines of thought that in Sect.~\ref{sec:azimuthal_perturbations} established  the lower bound  \eqref{eq:bound_from_below_az}, here we arrive at  
\begin{equation}
\label{eq:bound_from_below_r}
\delta^2\mathcal{F}[U]\geqq\int_0^{1}\left(\frac{\psi}{\gamma}-\frac{\gamma'^2}{\gamma}-\gamma''\right)U(u)^2\dd u
\end{equation}
and the following inequality suffices to render $\delta^2\mathcal{F}$ positive,
\begin{align}
\label{eq:inequality_r}
\frac{\psi}{\gamma}-\frac{\gamma'^2}{\gamma}-\gamma''&=\frac{4A^2u}{\sqrt{1-(Au)^2}\sqrt{1+(k_3-1)(Au)^2}}\left[(k_3-1)\sqrt{1-(Au)^2}\sqrt{1+(k_3-1)(Au)^2}\right.\nonumber\\
& \left. +\frac{1}{4}(Au)^2(k_3+11)(k_3-1)-k_3+2\right]\geqq 0,
\end{align}
which is indeed satisfied for all $u\in[0,1]$ and $(k_3,A)\in\conf$.

We thus conclude that the ET fields are locally stable, despite the violation of one Ericksen's inequality.\footnote{A violation nonetheless necessary for these fields to be equilibrium solutions.} Here we considered only the case $k_3\geqq1$; for completeness, the local stability of the ET fields has also been established  in the case $0<k_3\leqq1$ by use of a spectral method  \cite{paparini:thesis}. 

\section{Conclusions}\label{sec:conclusions}
In this article, we addressed the problem posed by the peculiar ground state exhibited by chromonic liquid crystals. These, unlikely ordinary nematic liquid crystals, when confined within capillary tubes with degenerate boundary conditions do not acquire the uniform alignment with the director oriented along the cylinder's axis, but develop a spontaneous twist, equally likely to have opposite chiralities. A commonly accepted explanation for such a behaviour is that these materials have a twist elastic constant $K_{22}$ smaller than the saddle-splay  constant $K_{24}$ of Frank's elastic theory, a fact which has been confirmed by a number of experiments. The problem with this explanation is that assuming $K_{22}<K_{24}$ violates one of Ericksen's inequalities, which guarantee that Frank's elastic free energy is bounded below.

Very recently, such a violation has also been investigated in \cite{long:violation}, which proposed that the pure (double) twist mode that would characterize the ground state of chromonics,\footnote{With only $T\neq0$.} being non-uniform and so unable to fill space \cite{virga:uniform}, prompts the excitation of other elastic modes whose positive cost counterbalances the divergence to negative infinity of the total free energy. Here, concerned as we were by tackling a variational problem with indefinite energy, we gave a different, but complementary explanation. We showed that the  boundary conditions prescribed on the capillary tubes are indeed responsible for taming the unboundedness of the energy and securing a solution to the variational problem in an admissible class of distortions with cylindrical symmetry.

The issue about the stability of such a solution then remained open, a question that was not idle to ask, given the wildness of the parent energy. To resolve this issue, we derived a general formula for the second variation of Frank's elastic free-energy functional and applied it to the study of the (local) stability of the twisted ground state of chromonics. We concluded that this is stable. 

The role played by boundary conditions in stabilizing the ground state of chromonics within the classical elastic theory of Frank poses at least two further questions. First, what is the most general class of anchoring conditions capable in rigid containers of preventing the energy from diverging to negative infinity? Second, and more subtly, would \emph{free} boundaries, such as the ones arising in the equilibrium problem for a chromonic drop surrounded by its isotropic melt, keep the energy bounded below?

We have not tackled the first question. As for the second, preliminary explorations \cite{paparini:thesis} have shown that the violation of Ericksen's inequalities for either the twist or splay constants may result in a paradoxical disintegration process, which poses a serious threat to the applicability of Frank's elastic theory to chromonics. More details on such a threat will be given in a forthcoming paper \cite{paparini:paradoxes}, where we shall also discuss the amendments to the classical elastic theory that may be needed for it to encompass chromonics safely.

\appendix
\section{Dynamical Analogy}\label{sec:analogy}
This Appendix is devoted to a dynamical interpretation of the free-energy functional $\mathcal{F}$ in \eqref{eq:free_bur}. In this interpretation, the first integral \eqref{eq:first_intergral_Burylov} will be regarded as a conservation law and the qualitative features of the equilibrium solution \eqref{eq:bur_solution} will be derived from a phase plane analysis. Of course, nothing will be added to what we already know about the ET fields, but we shall perhaps appreciate better their stance amid other equilibrium configurations with unbounded energy.

We transform functional \eqref{eq:free_bur} into a dynamical \emph{action} by introducing the artificial time  
\begin{equation}
	\label{eq:change_variables_dyn}
	t:=-\ln\rho.
\end{equation}
Thus, the axis of the cylinder $\body$ at $ \rho = 0 $ is approached as $t\to\infty$, while $t = 0$, the origin of time, corresponds to the surface of the lateral surface of $\body$. Correspondingly, the angle $\beta$ becomes a function of $t$, 
\begin{equation}
	\label{eq:b_func_dyn}
	b(t):=\beta\left(\mathrm{e}^{-\rho}\right),
\end{equation}
and $\mathcal{F}$ in \eqref{eq:free_bur} acquires the  form of an \emph{infinite horizon} action \cite{agrachev:smooth}, 
\begin{equation}
	\label{eq:free_bur_dyn0}
	\mathcal{F}[b]:= \int_0^{\infty}\left(\frac{\dot{b}^2}{2}+\Phi(b)\right) \dd t-\Phi_0(b(0)),
\end{equation}
where, as customary, a superimposed dot denotes differentiation with respect to $t$ and
\begin{equation}
	\label{eq:phi_dyn}
	\Phi(b):=\frac12\sin^2b(\cos^2b+k_3\sin^2b),
\end{equation}
\begin{equation}
	\label{eq:phi_0dyn}
	\Phi_0(b):=-\frac12(1-2k_{24})\sin^2b.
\end{equation}

The associated Lagrangian $\mathcal{L}$ is the classical sum of a kinetic energy and a potential,
\begin{equation}
	\label{eq:lagrangian}
	\mathcal{L}:=\frac12\dot b^2+\Phi(b),
\end{equation}
in the single Lagrangian variable $b$.
It is perhaps less typical of a classical dynamical system the \emph{initial-value} potential $\Phi_0$.  The \emph{orbits} of the system are all 
solutions to the equation of motion
\begin{equation}
	\label{eq:eq_motion}
	\frac{\dd}{\dd t}\frac{\partial \mathcal{L}}{\partial \dot b}-\frac{\partial \mathcal{L}}{\partial b}=0,
\end{equation}
which here reads simply as
\begin{equation}
	\label{eq:EL_bulk}
	\ddot b-\Phi'(b)=0.
\end{equation}
This equation is subject to both initial and asymptotic conditions which stem from requiring stationarity of the action at the beginning and at the end of time,\footnote{Alternatively, we may say that equations \eqref{eq:momentum_sources} prescribe the sources of momentum at both ends of the time horizon.}
\begin{subequations}\label{eq:momentum_sources}
	\begin{align}
\left(\frac{\partial \mathcal{L}}{\partial \dot b}+\frac{\partial \Phi_0(b)}{\partial b}\right)_{t=0}&=0, \label{eq:eq_motion_boundary}\\
	\lim_{t\to\infty}\frac{\partial\mathcal{L}}{\partial \dot b}&=0.	\label{eq:eq_motion_asymptotic}
	\end{align}
\end{subequations}
While the latter requires that
\begin{equation}\label{eq:asymptotic_condition}
\lim_{t\to\infty}\dot{b}(t)=0,	
\end{equation}	
the former identifies a locus in the phase plane $(b,\dot b)$ for all admissible initial conditions,
\begin{equation}
	\label{eq:natural_condition}
	\dot b=(1-2k_{24})\sin b\cos b.
\end{equation}

\begin{figure}[h]
	\centering
	\includegraphics[width=0.5\linewidth]{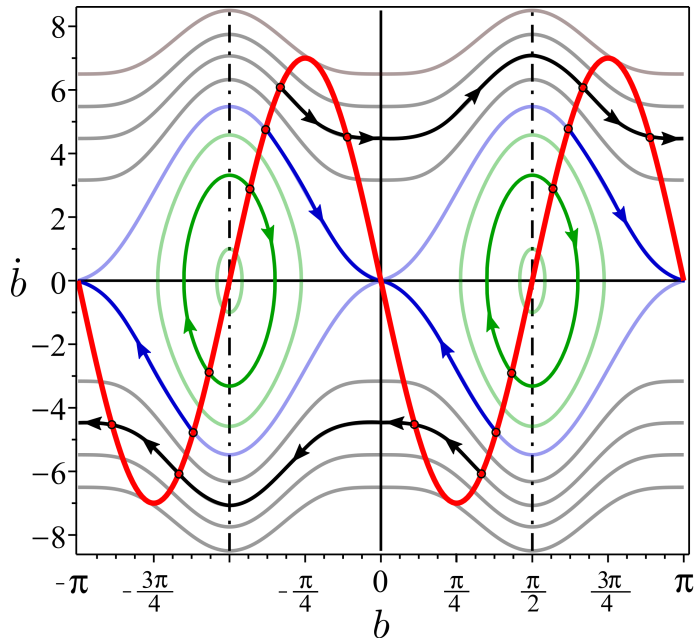}
	\caption{Phase portrait in the plane $(b,\dot b)$ of the dynamical system with total action in \eqref{eq:free_bur_dyn0}. Here $k_3=30$, $k_{24}=7.5$, and $c\geqq-30$. The red curve is the locus of allowed initial conditions; the separatrix is blue, while closed and open orbits are depicted in green and gray, respectively. Arrows indicate the direction of flow as time elapses. Three trajectories have been highlighted in darker green, darker blue, and black, starting from the initial condition: they correspond to the following values of the parameter that selects the orbits, $c=-20$, $c=0$, and $c=20$, respectively.}
	\label{fig:phase_portrait}
\end{figure}

A   \emph{conservation law} follows readily from \eqref{eq:EL_bulk},
\begin{equation}
	\label{eq:conservation_law}
	\frac12\dot b^2=\Phi(b)+\frac12 c,
\end{equation}
where $c$ is an arbitrary constant. Since $\Phi\leqq k_3/2$, it readily follows from \eqref{eq:conservation_law} that the admissible values of $c$ fall in the range $c\geqq-k_3$. The phase portrait of \eqref{eq:conservation_law} in the plane $(b,\dot b)$ is illustrated in Fig.~\ref{fig:phase_portrait} for $b\in[-\pi,\pi]$.
Orbits are selected by the values of $c$. For $c=-k_3$, the corresponding orbits collapse in the points $(\pm\pi/2,0)$. As $c$ increases, orbits are inflated and stay closed until $c=0$ (see green trajectories in Fig.~\ref{fig:phase_portrait}). This latter value corresponds to orbits that converge to the origin and to the points $(\pm\pi,0)$ (blue trajectories). The collectively act as a \emph{separatrix} in phase plane, bounding the domain of closed orbits. For $c>0$, orbits become unbounded (gray trajectories). 

This qualitative analysis, based on the conservation law \eqref{eq:conservation_law}, embraces all possible orbits. We are especially interested in those that start from the admissible locus \eqref{eq:natural_condition} in phase space and obey the asymptotic condition \eqref{eq:asymptotic_condition}. The former is represented by the red curve in Fig.~\ref{fig:phase_portrait}, while the latter requires the orbit to approach the $b$-axis as $t\to\infty$. Now, if $c<0$, the green trajectories are periodic orbits that keep crossing infinitely many times the $b$-axis, whereas if $c>0$ the gray trajectories are open orbits that never cross the $b$-axis. Only for $c=0$, that is, along the separatrix, can an orbit  approach the $b$-axis starting from an admissible initial condition. 

As shown in Fig.~\ref{fig:phase_portrait}, there is a critical value of $c$ corresponding to orbits tangent to the curve of admissible initial conditions,
\begin{equation}
	\label{eq:c_max}
	c_{\mathrm{max}}:=\frac{4k_{24}^2(k_{24}-1)^2}{k_3+4k_{24}(k_{24}-1)}.
\end{equation}
For $c>c_\mathrm{max}$, the open orbits of the system are inadmissible. For $0<c\leqq c_\mathrm{max}$ and $-k_3\leqq c<0$, the orbits obey \eqref{eq:eq_motion_boundary}, but violate \eqref{eq:eq_motion_asymptotic}; their total action $\mathcal{F}$ in \eqref{eq:free_bur_dyn0}, which by \eqref{eq:conservation_law} can be expressed as
\begin{equation}
	\label{eq:free_c_phi0}
	\mathcal{F}[b]=\int_0^{\infty}\left[2\Phi(b)+\frac{c}{2}\right]\dd t+ \Phi_0(b(0)),
\end{equation}
is easily seen to be unbounded.

Contrariwise, for $c=0$, the total action is finite and can be shown to equal the elastic free energy $\mathcal{F}_\mathrm{ET}$ computed in \eqref{eq:free_energy_ET_beta_1}. The rate at which time diverges as the origin in phase space is approached (on a the separatrix) can be estimated as 
\begin{equation}
	\label{eq:time_lapse_c0}
	t\approx-2\ln b\quad\text{as}\quad b\to0,
\end{equation} 
see \cite{paparini:thesis} for more quantitative details. 

\begin{figure}[h]
	\centering
	\begin{subfigure}[c]{0.45\linewidth}
		\centering
		\includegraphics[width=\linewidth]{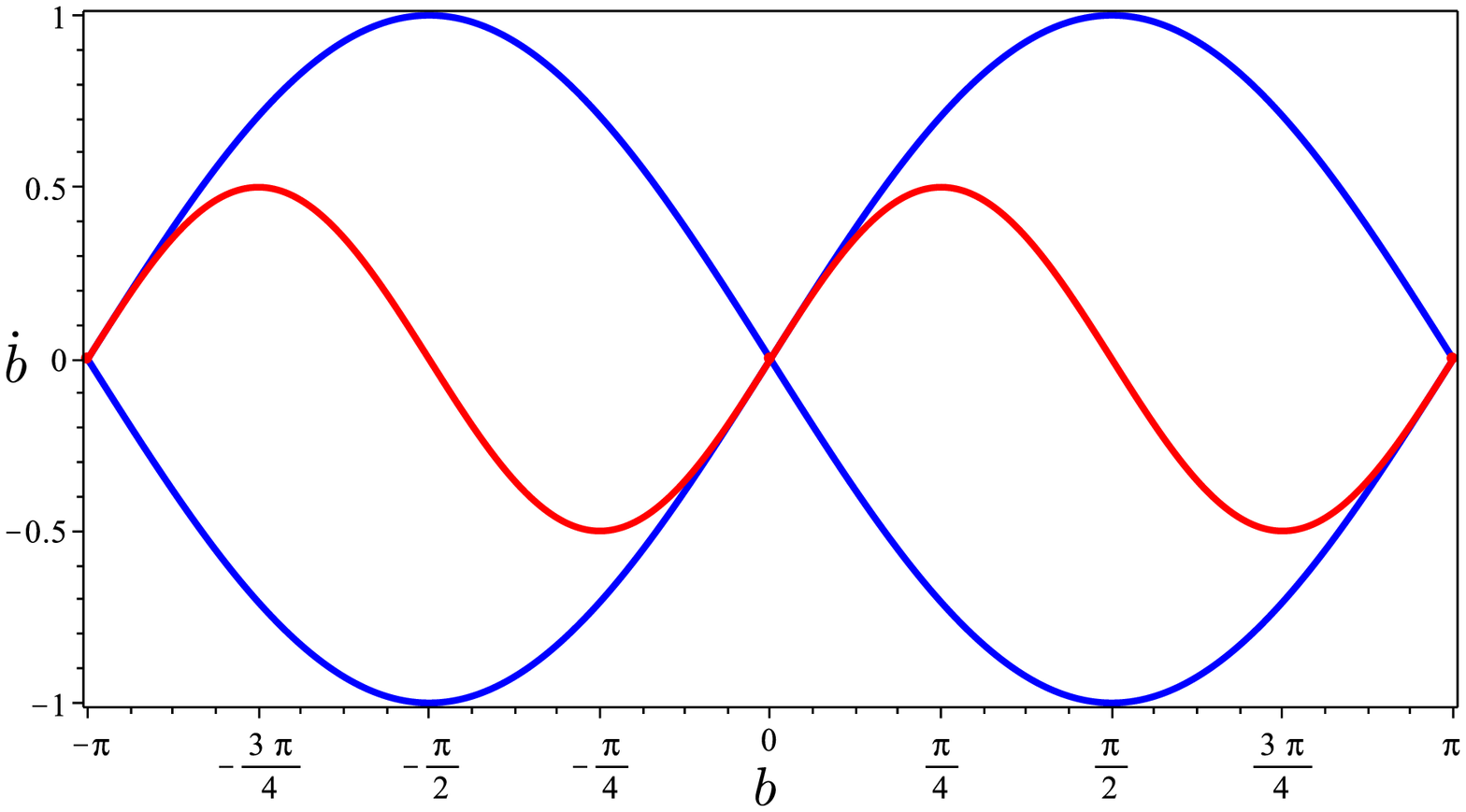}
		\caption{$k_{24}=0$}
		\label{fig:violation_a}
	\end{subfigure}
	\quad
	\begin{subfigure}{0.45\linewidth}
		\centering
		\includegraphics[width=\linewidth]{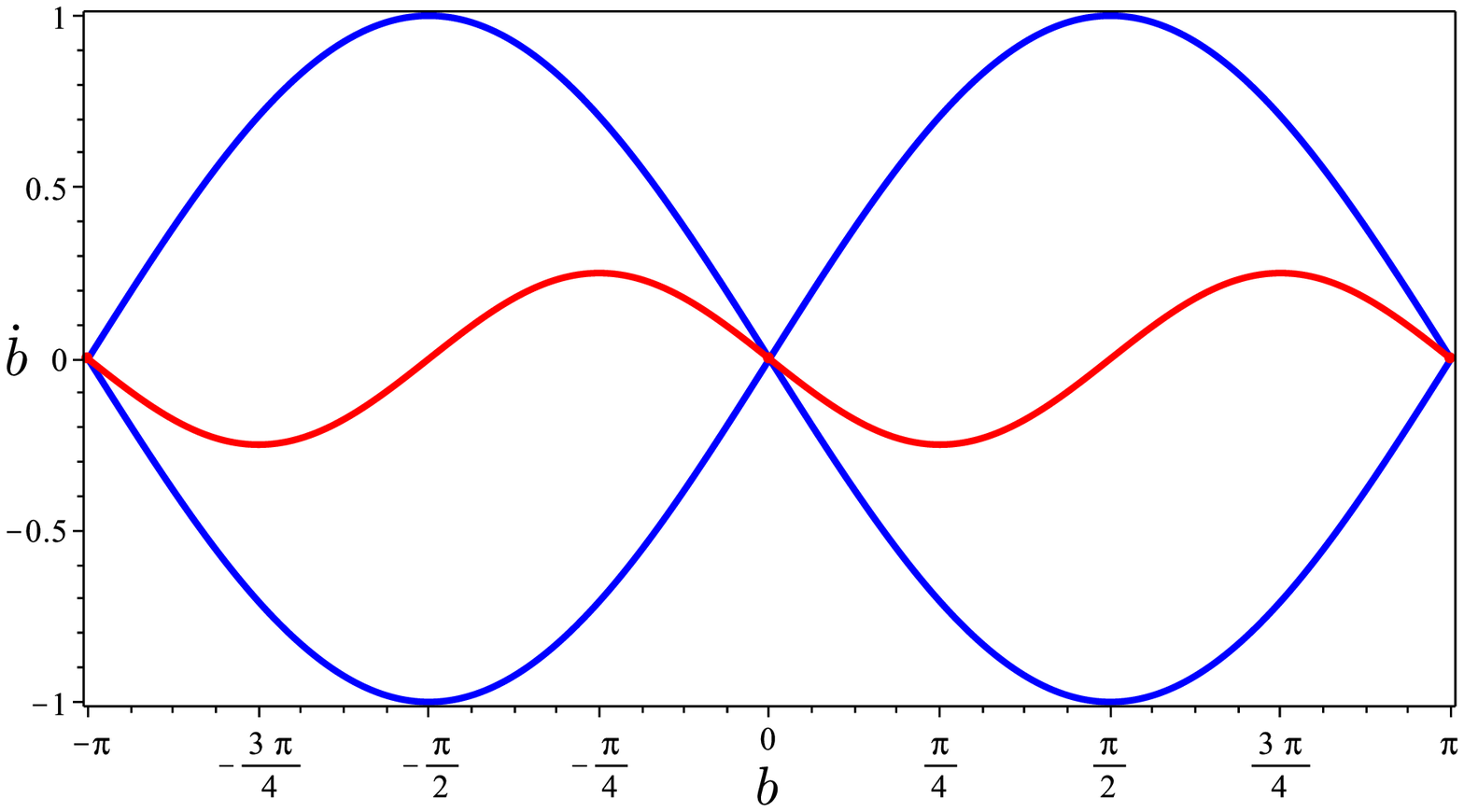}
	\caption{$k_{24}=\frac34$}
	\label{fig:violation_b}
	\end{subfigure}
\\
	\begin{subfigure}[c]{0.45\linewidth}
	\centering
	\includegraphics[width=\linewidth]{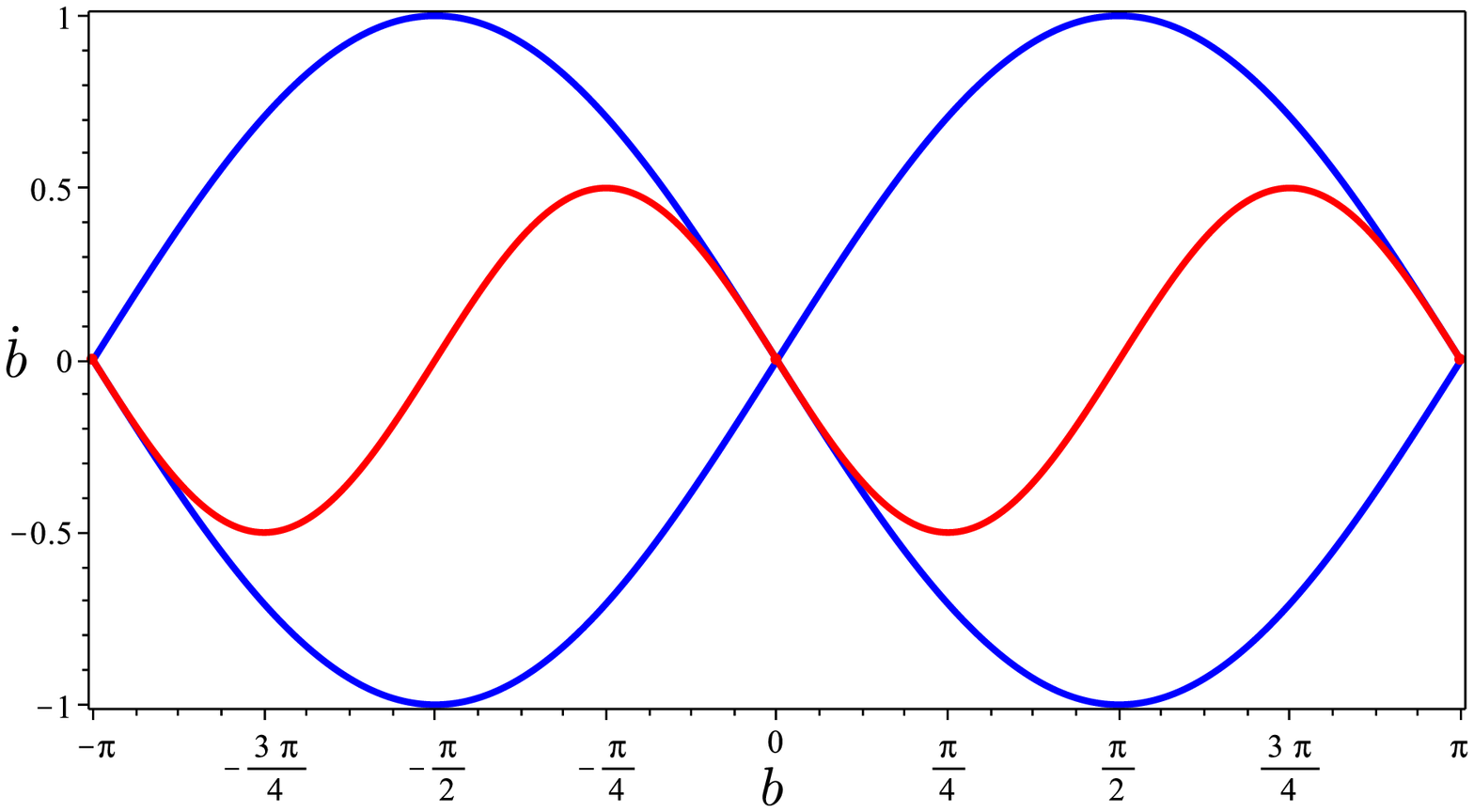}
	\caption{$k_{24}=1$}
	\label{fig:violation_c}
\end{subfigure}
\quad
\begin{subfigure}{0.45\linewidth}
	\centering
	\includegraphics[width=\linewidth]{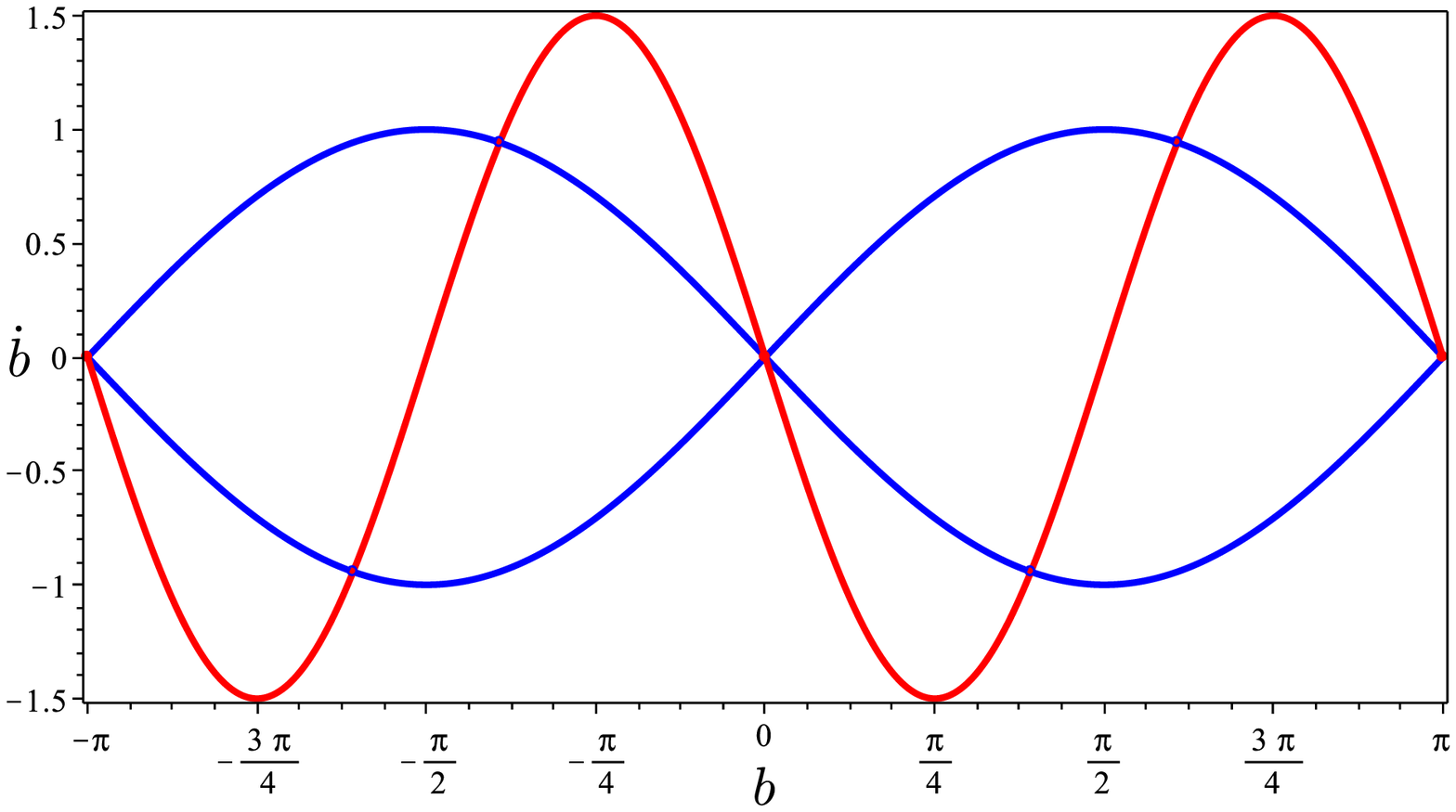}
	\caption{$k_{24}=2$}
	\label{fig:violation_d}
\end{subfigure}
	\caption{The separatrix (blue) and  the curve of allowed initial conditions (red) are depicted on the phase plane $(b,\dot b)$ for $k_3=1$ and several values of $k_{24}$. For $0\leqq k_{24}\leqq1$, no orbit is allowed with finite total action, as the curve of admissible initial conditions is fully enclosed by the separatrix.}
	\label{fig:violation}
\end{figure}

We have learned in Sect.~\ref{sec:ground} that the equilibrium solution $\beta_\mathrm{ET}$ in \eqref{eq:bur_solution} only exists for $k_{24}>1$. This also arises from the dynamical analogy studied here and is revealed by a geometric feature of the phase portrait. As can be shown by an asymptotic expansion of both  \eqref{eq:natural_condition} and \eqref{eq:conservation_law} near the origin, for $k_{24}<1$, the curve of admissible initial conditions lies inside the separatrix, and so there is no admissible orbit with finite total action. Figure~\ref{fig:violation} illustrates graphically the situation; in particular, for $k_{24}=0$ and $k_{24}=1$, the curve of admissible initial conditions is (internally) tangent to the separatrix.
  
\section{Useful Identities and Inequalities}\label{sec:identities}
We collect in this Appendix a number of identities and inequalities that are used in the main text.

For the mapping $t\mapsto\nt(\x)$ introduced in Sect.~\ref{sec:second_variation}, we compute the following derivatives with respect to the parameter $t$,
\begin{equation}
	\label{eq:trace_n}
	\frac{\dd}{\dd t}\tr\left(\nabla\nt\right)^2=\frac{\dd}{\dd t}\left(\I\cdot\nabla\nt^2\right)=\I\cdot\left(\nabla\dotnt\right)\nabla\nt+\I\cdot\nabla\nt\left(\nabla\dotnt\right)=2\tr(\nabla\dotnt\nabla\nt)
\end{equation}
and
\begin{equation}
	\label{eq:trace_n_second}
	\frac{\dd^2}{\dd t^2}\tr\left(\nabla\n(t)\right)^2=2\frac{\dd}{\dd t}\left(\I\cdot\nabla\dotnt\nabla\nt\right)=2\tr\left(\nabla\ddotnt\nabla\nt+(\nabla\dotnt)^2\right).
\end{equation}
By use of these in \eqref{eq:free_energy_density}, we readily obtain that 
\begin{align}
	\label{eq:first_derivative_W}
	\dot{W}_\mathrm{F}&:=\frac{\dd}{\dd t}\WF(\nt,\nabla\nt)=K_{11}\diver\nt\diver\dotnt+K_{22}(\nt\cdot\curl\nt)\left(\dotnt\cdot\curl\nt+\nt\cdot\curl\dotnt\right)\nonumber\\
	&+K_{33}\nt\times\curl\nt\cdot\left(\dotnt\times\curl\nt+\nt\times\curl\dotnt\right)+2K_{24}\left[\tr(\nabla\dotnt\nabla\nt)-\diver\nt\diver\dotnt\right]
\end{align}
and
\begin{align}
	\label{eq:second_derivative_W}
	\ddot{W}_\mathrm{F}:=&\frac{\dd^2}{\dd t^2}\WF(\nt,\nabla\nt)=K_{11}\Big\{(\diver\dotnt)^2+\diver\nt\diver\ddotnt\Big\}\nonumber \\
	&+K_{22}\Big\{\left(\dotnt\cdot\curl\nt+\nt\cdot\curl\dotnt\right)^2
	+\nt\cdot\curl\nt\left(\ddotnt\cdot\curl\nt+2\dotnt\cdot\curl\dotnt+\nt\cdot\curl\ddotnt\right)\Big\}\nonumber\\
	&+K_{33}\Big\{\left|\dotnt\times\curl\nt+\nt\times\curl\dotnt\right|^2+\nt\times\curl\nt\cdot\left(\ddotnt\times\curl\nt+2\dotnt\times\curl\dotnt+\nt\times\curl\ddotnt\right)\Big\}\nonumber\\
	&+2K_{24}\Big\{\tr\left(\nabla\ddotnt\nabla\nt\right)+\tr(\nabla\dotnt)^2-(\diver\dotnt)^2-\diver\nt\diver\ddotnt\Big\}.
\end{align}

For a director field $\n$ as in \eqref{eq:n_bur}, whose gradient has been computed in \eqref{eq:grad_B}, and for the perturbation field $\vv$ described in \eqref{eq:Burylov_perturbation}, the following computational ingredients were needed in the main text to arrive from the general formula for the second variation of Frank's elastic free energy in \eqref{eq:second_variation} at the special form needed when the unperturbed field in $\n=\nB$,
\begin{subequations}\label{eq:distortion_measures_nB_v}
	\begin{comment}
	\begin{equation}
		\nabla\nB=\frac{1}{R}\left(-\frac{\sin\beta_\mathrm{ET}}{\rho}\e_r\otimes\e_\vt+\cos\beta_\mathrm{ET}\beta_\mathrm{ET}'\e_\vt\otimes\e_r-\sin\beta_\mathrm{ET}\beta_\mathrm{ET}'\e_z\otimes\e_r\right),
	\end{equation}
\end{comment}
	\begin{equation}
		\diver\n=0,\label{eq:div_bur}
	\end{equation}
	\begin{equation}
		\n\cdot\curl\n=\frac{1}{R}\left(\beta'+\frac{\cos\beta\sin\beta}{\rho}\right),\label{eq:twist_bur}
	\end{equation}
	\begin{equation}
		\n\times\curl\n=\frac{1}{R}\left(\frac{\sin^2\beta}{\rho}\right)\e_r,\label{eq:bend_bur}
	\end{equation}
	\begin{equation}
		\tr\left(\nabla\n\right)^2=\frac{1}{R^2}\left(-\frac{2\cos\beta\sin\beta\beta'}{\rho}\right),\label{eq:saddle_splay_bur}
	\end{equation}
	\begin{align}
		\nabla\vv=\frac{1}{R}\bigg[& f'\e_r\otimes\e_r+\frac{g\cos\beta}{\rho}\e_r\otimes\e_{\vt}+\left(g\sin\beta\beta'-\cos\beta g'\right)\e_{\vt}\otimes\e_r\nonumber\\
		&+\frac{f}{\rho}\e_{\vt}\otimes\e_{\vt}+\left(\sin\beta g'+g\cos\beta\beta'\right)\e_z\otimes\e_r\bigg],\label{eq:grad_pert}
	\end{align}
	\begin{equation}
		\diver\vv=\frac{1}{R}\left(f'+\frac{f}{\rho}\right),\label{eq:div_pert}
	\end{equation}
	\begin{equation}
		\vv\cdot\curl\n=\frac{1}{R}\left(\frac{g\sin^2\beta}{\rho}\right),\label{eq:twist_n}
	\end{equation}
	\begin{equation}
		\n\cdot\curl\vv=\frac{1}{R}\left[-g'-\frac{g\cos^2\beta}{\rho}\right],\label{eq:twist_v}
	\end{equation}
	\begin{equation}
		\vv\cdot\curl\vv=\frac{1}{R}\left[g^2\left(\beta'-\frac{\sin\beta\cos\beta}{\rho}\right)\right],\label{eq:twist_pert}
	\end{equation}
	\begin{equation}
		\vv\times\curl\n=\frac{1}{R}\left[-g\left(\beta'+\frac{\sin\beta\cos\beta}{\rho}\right)\e_r-f\left(\frac{\sin\beta}{\rho}+\cos\beta\beta'\right)\e_\vt+f\sin\beta\beta'\e_z\right],\label{eq:bend_n}
	\end{equation}
	\begin{equation}
		\n\times\curl\vv=\frac{1}{R}\left[g\left(\beta'-\frac{\sin\beta\cos\beta}{\rho}\right)\e_r\right],\label{eq:bend_v}
	\end{equation}
	\begin{equation}
		\left(\vv\times\curl\vv\right)\cdot\e_r=\frac{1}{R}\left(gg'+\frac{g^2\cos^2\beta}{\rho}\right),\label{eq:bend_pert}
	\end{equation}
	\begin{equation}
		v^2=f^2+g^2,\label{eq:norm}
	\end{equation}
	\begin{equation}
		\nabla v^2=\frac{1}{R^2}\left(2ff'+2gg'\right),\label{eq:norm_grad}
	\end{equation}
	\begin{equation}
		\tr\left(\nabla\vv\right)^2=\frac{1}{R^2}\left[f'^2+\frac{f^2}{\rho^2}+\frac{2g\cos\beta}{\rho}\left(g\sin\beta\beta'-\cos\beta g'\right)\right],\label{eq:saddle_splay_bur}
	\end{equation}	
\end{subequations} 
where a prime denotes differentiation with respect to $\rho$.

A family of inequalities has played a central role in Sect.~\ref{sec:local}. They stem from the adaptation to our setting of a Lemma proved in  \cite{li1995:inequalities} (see Lemma\, 2.1).
\begin{lemma}\label{lemma:li1995}
	Let $[a,b]$ be an interval in $\mathbb{R}$ and let $\widetilde{U}=\widetilde{U}(u)\in AC[a,b]$.  If there are two functionals, $Q(u,t)$ and $G(u,t)$, satisfying the following conditions
	\begin{enumerate}
		\item $t$ is in the range of the function $\widetilde U=\widetilde U(u)$,
		\item $Q(u,\widetilde{U}(u))$ is integrable for $u\in[a,b]$,
		\item $G(u,\widetilde{U}(u))$ is absolutely continuous for $u\in[a,b]$,
	\end{enumerate}
	then 
	\begin{equation}
		\label{eq:energy_comparison}
		\int_a^b Q(u,\widetilde{U}(u))\widetilde{U}'(u)^2\dd u\geqq-\int_a^b\left[Q(u,\widetilde{U}(u))^{-1}G_{\widetilde{U}}^2+2G_u\right]\dd u+2\left[G(b,\widetilde{U}(b))-G(a,\widetilde{U}(a))\right],
	\end{equation}
	where $G_{\widetilde{U}}:=\partial G(u,\widetilde U)/\partial \widetilde U$ and $G_{u}:=\left[\partial G(u,t)/\partial u\right]|_{t=\widetilde U}$.
\end{lemma}

\begin{comment}
\section*{Data Availability Statement}
Data sharing not applicable to this article as no datasets were generated or analysed during the current study.

\bibliography{Chromonics}
\end{comment}

%

\end{document}